\documentclass[prd,reprint,amsmath,nofootinbib]{revtex4-1}
\pdfoutput=1

\usepackage{ifthen}
\usepackage{epsfig}
\usepackage{booktabs} 
\usepackage{natbib}
\usepackage{wasysym}
\usepackage{feynmp}
\usepackage{slashed} 

\usepackage{color}

\newcommand{\beqra}{\begin{eqnarray}}
\newcommand{\eeqra}{\end{eqnarray}}
\newcommand{\beq}{\begin{equation}}
\newcommand{\eeq}{\end{equation}}

\begin{document}

\title{Direct Detection of Dark Matter Bound to the Earth}

\author{Riccardo Catena}
\email{catena@chalmers.se}
\affiliation{Chalmers University of Technology, Department of Physics, SE-412 96 G\"oteborg, Sweden}
\author{Chris Kouvaris}
\email{kouvaris@cp3.sdu.dk}
\affiliation{CP$^3$-Origins, University of Southern Denmark, Campusvej 55, DK-5230 Odense, Denmark}

\begin{abstract}
We study the properties and direct detection prospects of an as of yet neglected population of dark matter (DM) particles moving in orbits gravitationally bound to the Earth.~This DM population is expected to form via scattering by nuclei in the Earth's interior.~We compute fluxes and nuclear recoil energy spectra expected at direct detection experiments for the new DM population considering detectors with and without directional sensitivity, and different types of target materials and DM-nucleon interactions.~DM particles bound to the Earth manifest as a prominent rise in the low-energy part of the observed nuclear recoil energy spectrum.~Ultra-low threshold energies of about 1 eV are needed to resolve this effect.~Its shape is independent of the DM-nucleus scattering cross-section normalisation.
\\[.1cm]
{\footnotesize  \it Preprint: CP3-Origins-2016-036 DNRF90}
\end{abstract}


\maketitle

\section{Introduction}
The detection of Milky Way dark matter (DM) particles is one of the most pressing research questions in Astroparticle Physics.~The experimental technique known as direct detection will play a crucial role in this context in the coming years~\cite{Baudis:2012ig}.~It searches for nuclear recoil events induced by the non-relativistic scattering of Milky Way DM particles in low-background detectors~\cite{Goodman:1984dc}.~The goal is to disentangle the expected DM signal, i.e.~a few nuclear recoil events per ton per year, from background events induced by environmental radioactivity, muon-induced neutrons or solar and atmospheric neutrinos~\cite{Lewin:1995rx}.~In order to achieve this goal, different experimental read-out strategies are currently under investigation, including the detection of scintillation light, athermal phonons, ionisation charge, and bubble nucleation~\cite{Undagoitia:2015gya}.~An alternative to background discrimination is the detection of an annual modulation in the observed rate of nuclear recoil events, which would allow to identify the DM origin of the observed signal unambiguously~\cite{Drukier:1986tm,Freese:1987wu,DAMA}.~The first ton-scale detectors for DM direct detection exploiting liquid Xenon or Argon are currently in a construction or commissioning stage~\cite{Baudis:2014naa}.~The first data release of XENON1T is for instance expected in 2017, with great expectations for groundbreaking discoveries~\cite{Aprile:2015uzo}.~At the same time, detectors with directional sensitivity, i.e.~designed to measure anisotropies in the distribution of nuclear recoil events, are currently in a research and development stage, and some first encouraging results have already been achieved~\cite{Mayet:2016zxu}. 

Low-threshold detectors are a priority in the design of DM direct detection experiments.~A first motivation for low-threshold detectors arises from models of light DM~\cite{Essig:2015cda}.~A DM particle of mass $m_\chi$ moving at a speed of $10^{-3}$ in natural units can deposit at most an energy $2\times 10^{-6} m_\chi^2 m_N/(m_\chi + m_N)^2$ in the scattering by nuclei of mass $m_N$.~Therefore, it is required a threshold energy of about 1 keV (1 eV) to detect a 1 GeV (1 MeV) DM particle in DM-nucleus elastic collisions. This can be somewhat improved by looking at inelastic channels~\cite{Kouvaris:2016afs}.~Currently none of the operating direct detection experiments has reached threshold energies of 1 eV yet. However, various strategies are under consideration, ranging from the initial proposal of Drukier and Stodolsky for the detection of neutrinos via neutral-current interactions~\cite{Drukier:1983gj} to more recent studies by where DM detection is achieved via  excitations in superfluid helium~\cite{Schutz:2016tid} or semiconductors~\cite{Hochberg:2016sqx}.

We have recently argued that low-threshold direct detection experiments are crucial for a second important reason~\cite{Catena:2016sfr}.~They would allow for the detection of an as of yet neglected population of DM particles gravitationally bound to the Earth, for which we have calculated the expected flux and induced event rate at detector.~This new population of DM particles would manifest in a direct detection experiment as a prominent spectral feature in the low-energy part of the observed nuclear recoil energy spectrum.~Such a population of bound DM particles can form if DM interacts with the nuclei in the Earth and scatters to orbits gravitationally bound to the planet, where it accumulates over the whole history of the solar system until the present time, when it is eventually detected.~The velocity distribution of this new population of DM particles peaks just below the Earth's escape velocity, and the induced nuclear recoil spectrum at detector is maximum for values of the DM particle mass close to the mass of abundant elements in the Earth, since in this mass range the probability of scattering to bound orbits is larger.

The literature on the capture of DM particles in orbits bound the solar system is considerable.~Most of these studies focus on the capture of DM particles by the Sun, and on the subsequent accumulation and annihilation of such particles at the Sun's centre, resulting in energetic neutrinos observable on Earth,~e.g.~\cite{Press:1985ug,1985ApJ...299.1001K,Peter:2009mi,Peter:2009mm}.~The direct detection of DM particles from orbits bound to the Sun is studied in~\cite{Damour:1998rh,Damour:1998vg}.~It is found that the expected rate of nuclear recoils is small due to the large Earth to Sun distance.~The capture of DM particles in orbits bound to the Earth is investigated in, e.g.~\cite{Freese:1985qw,Krauss:1985aaa,Gould:1987ir,Lundberg:2004dn}.~Most of the works on this topic focus on the neutrino signal produced by DM annihilation at the Earth's centre.~To the best of our knowledge, the direct detection of DM particles bound to the Earth is addressed in two articles only, besides our recent publication~\cite{Catena:2016sfr}.~In the pioneering work by Gould et al.~\cite{Gould:1988eq}, the direct detection of DM particles bound to the Earth is studied assuming a modified isothermal velocity distribution for DM.~This study carefully accounts for various effects related to the Sun's gravitational potential, but focuses on standard spin-independent dark matter-nucleon interactions only.~In a subsequent publication~\cite{Collar:1998ka}, an explicit expression for the velocity distribution at the Earth's surface of DM particles in orbits bound to the planet is found.~Our work~\cite{Catena:2016sfr} extends these first investigations by considering a broader set of dark matter-nucleon interactions, a refined chemical composition for the Earth, and detectors with and without directional sensitivity.~In the present study, we further extend the results presented in~\cite{Catena:2016sfr} by providing significantly more general expressions for fluxes and rates now valid for arbitrary dark matter-nucleon interactions, and considering different target materials for the assumed terrestrial detectors.

This paper is organised as follows.~In Sec.~\ref{sec:capture} we will review and significantly extend the calculations presented in Ref.~\cite{Catena:2016sfr}, providing all details needed to compute the flux of DM particles bound to the Earth potentially observable in a terrestrial detector.~In Sec.~\ref{sec:detection} we will convert this flux into a rate of nuclear recoil events, considering both non-directional and directional detectors, and expressing all equations in terms of general DM-nucleus scattering cross-sections.~In Sec.~\ref{sec:results} we will numerically evaluate the main equations previously derived and discuss how our conclusions depend on assumptions regarding the direct detection of DM particles bound to the Earth.~Finally, we will conclude in Sec.~\ref{sec:conclusion}.

\section{DM capture by the Earth}
\label{sec:capture}
The capture of DM particles by stellar objects and the Earth has been studied extensively in the past~\cite{Press:1985ug,Gould:1987ir}.~In particular, the capture of DM in the Sun and its subsequent distribution in bound elliptical orbits has been studied both analytically~\cite{Damour:1998rh,Damour:1998vg} and numerically~\cite{Peter:2009mi,Peter:2009mm}. Here we focus on DM capture by the Earth. The key point for the DM capture is that the particle should scatter underground to velocities that are below the escape velocity of that particular point of the Earth, thus leading to a gravitational bound orbit. 

Let us review the capture rate of halo DM particles to gravitationally bound orbits in the Earth after a scattering with a nucleus inside the Earth starting from first principles. Let us assume that the DM particle density inside the Earth at the  scattering point just before the scattering takes place is 
\begin{equation}
dn_{\chi}=f (\vec{x},\vec{v}) d^3x d^3v,
\end{equation}
where $f(\vec{x},\vec{v})$ is the DM distribution right before the collision (at position $\vec{x}$ with velocity $\vec{v}$). The number of DM scatterings per time per center of mass solid angle $d\Omega$ that takes place within an infinitesimal volume $d^3x$ inside the Earth with nuclei of the element $A$ of density $n_A(\bf x)$ is given by
\begin{equation}
\label{ee1}
d\dot{N}_A=d^3x n_A(\vec{x}) d^3v f(\vec{x},\vec{v})v\frac{d\sigma_A}{dE_R} dE_R,
\end{equation}
where $d\sigma_A/dE_R$ is the differential cross section per recoil energy $E_R$.
 Not all scatterings lead to capture. The capture condition for a scattering is for the particle to lose energy larger than the kinetic energy it had asymptotically far away from the Earth. 
The energy before the collision (i.e. kinetic plus potential one) is
\begin{equation}
\label{eb}
E_{\text{before}}=\frac{1}{2}m_{\chi} (v^2-v_{\text{esc}}^2(r))=\frac{1}{2}m_{\chi}v_{\infty}^2,
\end{equation}
where $v_{\text{esc}}(r)$ is the escape velocity from the Earth at a radius $r$ from the center of the Earth (i.e. at the place of the scattering) and $v_{\infty}$ 
is the velocity of the particle at an asymptotically far away distance from the Earth. The total energy after the collision must be negative in order for the DM particle to remain in a bound orbit around the Earth. Its value is
\begin{equation}
\label{ef}
 E_{\text{after}}=\frac{1}{2}m_{\chi} (v'^2-v_{\text{esc}}^2(r))=-\frac{G m_{\chi} M_{\oplus}}{2a}\equiv -\frac{1}{2}m_{\chi}\alpha,
\end{equation}
where $a$ is the major semi-axis of the elliptical orbit after the collision and $\alpha$ is defined as $\alpha \equiv G M_{\oplus}/a$ ($G$ being the gravitational constant and $M_{\oplus}$ the mass of the Earth). Using Eqs.~(\ref{eb}) and (\ref{ef}) we get the energy transfer $E_R$ 
\begin{equation}
\label{QQ}
E_R=\frac{1}{2}m_{\chi}(v^2-v_{\text{esc}}^2(r)+\alpha)=\frac{1}{2}m_{\chi} (v_{\infty}^2+\alpha).
\end{equation}
  Eq.~(\ref{QQ}) gives $dE_R=(1/2)m_{\chi} d\alpha$ and 
Eq.~(\ref{ee1}) now reads
\begin{equation}
\label{eq_m1}
d\dot{N}_A=\frac{1}{2}d^3x n_A(\vec{x}) d^3v f(\vec{x},\vec{v}) v \frac{d\sigma_A}{dE_R} m_{\chi} d\alpha \Theta_{\alpha},
\end{equation}
where $\Theta_{\alpha}$ represents a step function that enforces the kinematic constraint $E_R \leq \beta_+^A E_{kb}$ where $E_{kb}=(1/2)m_{\chi}v^2$ is the kinetic energy before the collision. We define  
\begin{equation}
\beta_{\pm}^A=\frac{4 m_{\chi}m_A}{(m_{\chi}\pm m_A)^2}.
\end{equation}
Using Eq.~(\ref{QQ}) the above condition can be written as 
$2E_R/m_{\chi}=v_{\infty}^2+\alpha \leq \beta_+^A v^2=\beta_+^A (v_{\infty}^2+v_{\text{esc}}^2(r))$.  Since $1/\beta_+^A- 1/\beta_-^A=1$ the above constraint can be rewritten as 
\begin{equation}
\Theta_{\alpha} \equiv \Theta \left [\beta_-^A \left (v_{\text{esc}}^2(r)-\frac{\alpha}{\beta_+^A} \right ) -v_{\infty}^2 \right ],
\end{equation}
where it is understood that the step function $\Theta(x)=1$ if $x\geq 0$ or 0 otherwise. Since the Earth is moving with respect to the rest frame of the DM halo, the flux of incoming particles is not going to be isotropic. This will also infuence the distribution of elliptical orbits for the captured DM particles. However merely due to the rotation of the Earth around its own axis, we expect that the distribution of the elliptical orbits will be to good approximation isotropic. For asymptotically far away distances from the Earth we use a Maxwell-Boltzmann distribution
\beq
\label{MB}
f_{\infty}(\vec{v}_{\infty})=\frac{n_{\chi}}{\pi^{3/2} v_0^3}\exp \left ( -\frac{(\vec{v}_{\infty}+\vec{v}_e)^2}{v_0^2} \right ),
\eeq
where $v_0=220$~km~s$^{-1}$ is the local standard of rest, $v_e=232$~km~s$^{-1}$ is the Earth velocity in the galactic rest frame, and $n_\chi$ the DM number density in the Earth's neighborhood.
 Liouville's theorem states that the distribution function remains constant along the trajectory of a particle, i.e. $f(\vec{x},\vec{v})=f_{\infty}[v_{\infty}(\vec{x},\vec{v})]$ where $f_{\infty}$ is the DM distribution far away from the Earth and $v_{\infty}^2=v^2-v_{\text{esc}}^2(r)$.  Taking the angular average of $f(\vec{x},\vec{v})$ defined as $\int f(\vec{x},\vec{v}) d^3v=4\pi\int v^2  \bar{f}(r,v) dv$ we get
\begin{align}
\label{fffu}
\bar{f}(v)  dv=\frac{n_\chi}{4\pi^{3/2}v_E v_0\sqrt{v^2-v_1^2}}\left(e^{-\frac{v_{-}^2}{v_0^2}}-e^{-\frac{v_{+}^2}{v_0^2}}\right)  dv\,,
\end{align}
where $v_{\pm}=\sqrt{v^2-v_1^2}\pm v_e$.  Note that we have dropped the variable $r$ from $\bar{f}$. The escape velocity of the Earth varies from 15~km~s$^{-1}$ at the Earth's centre to 11.2~km~s$^{-1}$ at the Earth's surface. Since the variation is small,  we simplified our calculation, by setting the escape velocity to its surface value $v_1=$11.2~km~s$^{-1}$. This makes $\bar{f}(r,v)$  independent of $r$ (leading to  Eq.~(\ref{fffu})). 

Upon making the isotropic approximation, we can simplify further Eq.~(\ref{eq_m1}). The specific angular momentum of the particle after the collision is $J=rv \sin\theta$ where $r$ is the distance from the center of the Earth, $v$ the velocity after the collision and $\theta$ the angle subtended by $\vec{r}$ and $\vec{v}$. Since we assume that $\cos\theta$ is uniformly distributed, and $J^2=J_{\text{max}}^2(1-\cos^2\theta)$, the distribution rewritten in terms of $J^2$ is $d\cos\theta = dJ^2/(2J_{\text{max}}^2\sqrt{1-J^2/J_{\text{max}}^2})$ where $J_{\text{max}}=rv$ is the maximum possible specific angular momentum after the collision. Within this approximation we can now rewrite Eq.~(\ref{eq_m1}) as
\begin{align}
\label{eq_m2}
d\dot{N}_A&=\pi d^3x n_A(r) v^3dv \bar{f}(v) \frac{d\sigma_A}{dE_R} m_{\chi} \left ( J_{\text{max}}^2 \sqrt{1-\frac{J^2}{J_{\text{max}}^2}} \right )^{-1} \\ \nonumber
&\times d\alpha dJ^2 \Theta_{\alpha} \Theta_J,
\end{align}
where $\Theta_J=\Theta(J_{\text{max}}-J)$ is a step function enforcing $J\leq J_{\text{max}}$. With the use of Eq.~(\ref{ef}), $J_{\text{max}}=r(v_{\text{esc}}^2(r)-\alpha)^{1/2}$. One can easily check that in the case of spin-independent interactions where
\beq
\label{SIcs}
\frac{d\sigma_A}{dE_R}=\frac{m_A\sigma_A}{2\mu_A^2 v^2}F_A^2(E_R),
\eeq
where $\mu_A$ is the DM-nucleus reduced mass, Eq.~(\ref{eq_m2}) becomes the one  derived in~\cite{Damour:1998vg}
\begin{align}
\label{dNdot}
 d\dot{N}_{A} = \frac{2\pi \sigma_Av \bar{f}(r,v)n_A(r) }{J_{\rm max}^2 \beta_{+}^A} &\left(1-\frac{J^2}{J^2_{\rm max}}\right)^{-1/2} F_A^2(E_R)\nonumber \\
&\times  \Theta_\alpha \Theta_{J} \left( d^3{\bf x}\,dv\right)\, d\alpha \, dJ^2 \,.
\end{align}
The form factor $F_A^2(E_R)$ accounts for the loss of coherence and it is  usually approximated by 
\begin{equation}
F_A^2(E_R)=\exp (-E_R/Q_A),
\end{equation}
where $E_R$ is the energy transferred during the collision and $Q_A=3/(2 m_A R_A^2)$, $m_A$ being the  nucleus mass and $R_A=10^{-13}\text{cm} \left 
[ 0.3 +0.91 \left (\frac{m_A}{\text{GeV}} \right )^{1/3} \right ]$ the radius of the nucleus. In this paper since we will present results for different types of DM-nuclei interactions, we will use Eq.~(\ref{eq_m2}) which can be used for any generic interaction and form factor.

Eq.~(\ref{eq_m2}) can be written in a more convenient form in terms of new more useful variables for the purposes of this study. Instead of using $J^2$ and $\alpha$, we will use the perihelion (minimum distance of the elliptical orbit to the center of the Earth) $r_m$ and the ellipticity of the orbit $e$. Recall that the semi-major axis for an ellipse is $a=r_m/(1-e)$ and consequently $\alpha=GM_{\oplus}(1-e)/r_m$. Note also that $J^2=r_m^2(v_1^2-\alpha)$.
From these two expressions we can calculate the Jacobian and get
\beq
dJ^2d\alpha=2 G M_{\oplus} \left (v_1^2-\frac{GM_{\oplus}(1-e)}{r_m} \right )de dr_m.
\eeq
Eq.~(\ref{eq_m2}) can be written in terms of the new variables $r_m$ and $e$ as
\begin{align}
\label{eq_m3}
d\dot{N}_A&=2\pi G M_{\oplus} d^3x n_A(r) v^3dv \bar{f}(v) \frac{d\sigma_A}{dE_R} m_{\chi} \\ \nonumber
&\times \left ( r^2 \sqrt{1-\frac{r_m^2}{r^2}} \right )^{-1}  \Theta_{r_m} \Theta_e dr_m de.
\end{align}
The condition $J=r_{m}(v_1^2-\alpha)^{1/2}\le J_{\rm max}$ imposed by $\Theta_J$ becomes $\Theta_{r_m}\equiv\Theta(r-r_m)$ and $\Theta_e$ is $\Theta_\alpha$ having subsituted $\alpha=GM_{\oplus}(1-e)/r_m$. Recall that the semi-major axis $a=r_m/(1-e)$. For the typical spin-independent DM-nucleus 
cross section of Eq.~(\ref{SIcs}), Eq.~(\ref{eq_m3}) takes the form provided in~\cite{Catena:2016sfr}
 \begin{align}
 d\dot{N}_A= 4\pi G M_{\oplus} \frac{\sigma_A v f(v) n_{A}(r)}{r^2\beta_{+}^A}
\left(1-\frac{r_m^2}{r^2} \right)^{-1/2}F^2_A(E_R)  \nonumber \\
\times\Theta_{r_{m}}\Theta_e \left( d^3{\bf x}\,dv\right)\, d e \,dr_m \,.
\end{align}
Since we  consider generic DM-nuclei interactions, we are going to use the more generic form of Eq.~(\ref{eq_m3}).

~Eq.~(\ref{eq_m3}) should be summed over all elements abundant in the Earth. In practice we take into account the most abundant elements, i.e. $^{16}$O, $^{28}$Si, $^{24}$Mg, $^{56}$Fe, $^{40}$Ca, $^{23}$Na, $^{32}$S, $^{59}$Ni, and $^{27}$Al assuming the standard composition and density profile of chemical elements in the Earth $n_A(r)$ provided in~\cite{Gondolo:2004sc}. Integrating Eq.~(\ref{eq_m3}) over $d^3{\bf x}\, dv$ and summing over elements gives
\begin{align}
\label{dNdot3}
& d \dot{N} = 8\pi^2 G M_{\oplus} m_{\chi} \sum_A  K_A(r_m,e) \nonumber \\ \times&\int_{r_m}^{R_{\oplus}} d r\, n_A(r) 
\left(1-\frac{r_m^2}{r^2} \right)^{-1/2}  d e \,{\rm d}r_m \equiv g(r_m,e)  de \, dr_m.
\end{align}
Eq.~(\ref{dNdot3}) gives the rate  of  accumulation of trapped DM particles into bound elliptical orbits of ellipticity within $[e,e+{\rm d}e]$, and perihelion within $[r_m,r_m+{\rm d}r_m]$.~In the derivation of Eq.~(\ref{dNdot3}), we have assumed spherical symmetry, i.e. ${\rm d}^3{\bf x}=4\pi r^2 {\rm d}r$. $K_A(r_m,e)$ is defined as
\begin{align}
\label{eq:K}
K_A(r_m,e) \equiv  \int_{v_1}^{v_2}  dv \,v^3 \bar{f}(v)\frac{d\sigma_A}{dE_R}.
\end{align}
 The upper limit $v_2$ comes from the step function $\Theta_e$ and it given by
\beq
\label{v2}
v_2=\sqrt{(1+\beta_{-}^A) v_1^2 - \frac{G M_\oplus}{r_m}(1-e)\frac{\beta_{-}^A}{\beta_{+}^A}} \,.
\eeq
The lower limit of intergration is obviously the escape velocity $v_1$ since a DM particle with zero speed at asymptotic far distances from the Earth, will acquire $v_1$ once it reaches the Earth. $d\sigma_A/dE_R$ depends generally on $E_R$ (either explicitly or via the form factor $F_A^2(E_R)$. In such a case 
\beq
E_R=(1/2)m_{\chi} \left (v^2-v_1^2+\frac{GM_{\oplus}(1-e)}{r_m} \right)
\eeq
is the energy loss in the collision that must be used in the evaluation of $K_A(r_m,e)$.

\section{Recoil Energy Spectrum of Bound Dark Matter}
\label{sec:detection}
In order to estimate the rate of events of bound DM particles scattering off a detector, we need to estimate the probability of  DM particles that follow a specific elliptic orbit to scatter off the detector as well as the number of bound DM particles per  specific elliptical orbit. To simplify our estimate, we are going to consider DM particles that have scattered in the Earth once in order to get captured and a second time in the detector creating a recoil signal. Multiple scatterings 
that take place underground diminish further the kinetic energy of the DM particle leading to recoil energies that are practically below any experimental threshold. 
 Therefore within this approximation, we estimate the number of DM particles that can accumulate in different orbits and have scattered only once. 
We can now estimate the number of periods $N$ required for a bound DM particle to scatter for a second time  
\begin{align}
\label{NN}
N=\left (\sum_A \int_0^{\theta_1} n_A(r) \sigma_A \xi(r_m,e) d\theta \right)^{-1},
\end{align}

where $\xi(r_m,e)d\theta$ is an infinitesimal  path along the elliptic trajectory of the orbit. The length of the path that a DM particle travels underground is
\beq 
\int d\ell=2\int_0^{\theta_1}d\theta \sqrt{\left( \frac{dr}{d\theta} \right )^2+r^2}\equiv \int_0^{\theta_1} \xi(r_m,e) d\theta.
\eeq 
Using the parametric equation for the elliptic orbit
\beq
\label{elip}
\frac{P}{r}=1+e \cos\theta,
\eeq
where $P$ is a constant, $e$ the ellipticity of the orbit and $\theta$ the angle subtented from a point of the orbit with distance $r$ from the center and the perihelion, it is easily found that 
 \beq \xi(r_m,e)= 2 r_m (1+e)\sqrt{1+e^2+2e\cos\theta}/(1+e\cos\theta)^2.
\eeq 
 The limit of integration $\theta_1$ is given by \beq \cos\theta_1 = \frac{r_m}{R_\oplus} \frac{(1+e)}{e} - \frac{1}{e}\eeq and corresponds to the angle subtended by the perihelion and the point where the orbit crosses the Earth ($r =R_\oplus$) from the Earth's center. It can be found by setting $r=R_{\oplus}$ and solve for $\theta$ in Eq.~(\ref{elip})
The condition $-1<\cos\theta_1<1$ implies that
\beq
\frac{1-e}{1+e} \le \frac{r_m}{R_\oplus} \le 1\,.
\eeq
For a given orbit, the time $T(r_m,e)$ a DM particle can  spend without scattering for a second time until today is on average
\begin{align}
\label{eq:Gamma}
T(r_m,e)\equiv \min [ N \times \tau(r_m,e), \tau_\oplus] \,,
\end{align}
where $\tau_\oplus\simeq 4.5\times 10^9$ years is the age of the Earth and
\beq \tau(r_m,e)= \sqrt{\frac{4\pi^2}{G M_\oplus} \frac{r_m^3}{(1-e)^3}}
\eeq is the period of the elliptical bound orbit.~We will refer to $T$ as accumulation time.

\subsection{Non-Directional Detectors}
The differential event rate in a non-directional detector for a given orbit characterized by $r_m$ and $e$ is
\begin{align}
\frac{dR_{r_m,e}}{dE_R}=N_T \frac{d\sigma_N}{dE_R} \mathcal{F}=N_T \frac{d\sigma_N}{dE_R}\frac{{\rm d} \dot{N}}{4 \pi l_c^2}\frac{2T(r_m,e)}{\tau (r_m,e)}, \label{difrate}
\end{align}
where $N_T$ is the number of target nuclei in the detector. $\mathcal{F}$ is the flux of bound DM particles in orbits of perihelion $r_m$ and ellipticity $e$ crossing the detector. The flux is equal to 
 the rate $d \dot{N}$ with which a particular orbit is populated (see Eq.~(\ref{dNdot3})) multiplied by the time $T(r_m,e)$ this orbit can accumulate DM particles divided by $\tau (r_m, e)/2$ since during each period of the orbit the DM particle crosses the Earth twice, divided by $4\pi \ell_c^2$ ($\ell_c$ being the distance between the detector and the center of the Earth). 
We have assumed that the elliptical orbits cross the surface of the Earth isotropically, i.e. there are no bound DM particles crossing a particular patch of the Earth's surface with a higher rate than another patch. This gives the factor $4\pi \ell_c^2$.
Since generically $d\sigma_N/dE_R$ depends on the DM particle velocity, it is needed to know the velocity before the scattering with the detector.
It is completely determined by $r_m$ and $e$ and can be easily shown to be 
\begin{align}
\label{vv}
v=\sqrt{2 G M_{\oplus} \left (\frac{1}{r}-\frac{1-e}{2 r_m} \right ) }.
\end{align}
with $r=\ell_c$.~Note that $d\sigma_N/dE_R$ refers to DM scattering off a detector nucleus and it should not be confused with   $d\sigma_A/dE_R$ that was the scattering that lead to the capture of DM by a random underground nucleus.

Combining Eqs.~(\ref{dNdot3}), (\ref{eq:Gamma}) and (\ref{difrate}) we obtain the differential rate of events
\begin{align}
\label{eq:rec}
\frac{dR}{dE_R}= \frac{N_T}{2 \pi \ell_c^2} \int_0^1 \int_{\frac{1-e}{1+e}R_\oplus}^{R_\oplus}de  dr_m g(r_m,e)\frac{d\sigma_N}{dE_R} \frac{T(r_m,e)}{\tau(r_m,e)} dr_m de.
\end{align}
We stress again that  in general $d\sigma_N/dE_R$ depends on $v$, and $v$ should be evaluated at the value given by Eq.~(\ref{vv}). Eq.~(\ref{eq:rec}) represents the main equation that gives the event rate in non-directional detectors. If one assumes spin-independent interactions (Eq.~(\ref{SIcs})), the spectrum recoil becomes 
\begin{align}
\label{eq:rec2}
\frac{dR}{dE_R}= \kappa & \int_0^1 \int_{\frac{1-e}{1+e}R_\oplus}^{R_\oplus} \frac{g(r_m,e)}{v^2}\frac{T(r_m,e)}{\tau(r_m,e)} dr_m de,
\end{align}
where
$
\kappa = N_T m_N \sigma_n A_N^2 F^2(E_R)/(4 \pi \ell_c^2 \mu_N^2)$. 

Eq.~(\ref{eq:rec}) must be contrasted to the recoil events coming from direct halo DM scatterings off nuclei targets in the detectors. The rate is as usually given by
\beq
\frac{dR}{dE_R}=N_T n_{\chi}\int_{v_{\text{min}}}^{v_{\text{esc}}+v_e}\frac{d\sigma_N}{dE_R}f(v)v d^3v,
\label{eq:halo}
\eeq
where $n_{\chi}$ is the local DM density in the Earth, and 
\beq 
\label{vmin}
v_{\text{min}}=\sqrt{m_NE_R/(2 \mu_N^2)}
\eeq
 is the minimum velocity that can produce nuclear recoil of energy $E_R$. For $f(v)$ we use the usual Maxwell-Boltzmann of Eq.~(\ref{MB}) with $v_{\text{esc}}$ and $v_e$ being the escape velocity of the Galaxy and the velocity of the Earth in the rest frame of the Galaxy respectively.\\

\subsection{Directional Detectors}
We also study  the spectrum of  bound DM scattering off directional detectors. By choosing an appropriate recoil direction, directional detectors have the advantage of minimizing the rate of events coming from the halo DM particles. Pointing the cone of detection along with the DM wind, one looks at particles that have velocities $\vec{v}-\vec{v}_e$. This leads to overall smaller  particle fluxes and consequently to smaller rate of events. On the contrary this choice does not affect the rate of events of bound DM particles. 
In particular we will consider the spectrum of recoils coming from a direction perpendicular to the vector that connects the center of the Earth with the detector.  We have found that such horizontal directions can give an enhancement in the bound/halo ratio of DM events in the detector. Generically the directional rate for energy recoil $E_R$ and recoil direction within the solid angle $d\Omega_q$ is 
\beq 
\label{directR}
\frac{dR}{dE_Rd\Omega_q}=N_T\int\frac{d\sigma}{dE_Rd\Omega_q}d\Phi,
\eeq where $d\Phi$ is the flux of particles arriving at the detector. For a generic DM-nucleus interaction, the cross section per nuclear recoil energy per recoil solid angle is 
\beq
\label{dir_cross}
\frac{d\sigma_N}{dE_R d\Omega_q}=\frac{d\sigma_N}{dE_R}\frac{1}{2 \pi}\delta \left ( \cos\theta_q-\frac{v_{\text{min}}}{v} \right ),
\eeq
where $\theta_q$ is the angle between the nuclear recoil and the initial DM velocity and $v_{\text{min}}$ is given by Eq.~(\ref{vmin}).
Eq.~(\ref{directR}) can be rewritten with  the  help of (\ref{dir_cross}) as
\begin{widetext}
\beq
\label{dir_rate}
\frac{dR}{dE_Rd\Omega_q}=\frac{N_T}{2 \pi \delta \ell_c^2} \int \frac{d\sigma_N}{dE_R}g(r_m,e)\frac{T(r_m,e)}{\tau(r_m,e)}\delta \left ( \cos\theta_q -\frac{v_{\text{min}}}{v} \right )dr_m de \frac{d\cos\theta d\phi}{4 \pi}\frac{d\omega}{2 \pi}.
\eeq
\end{widetext}
Eq.~(\ref{dir_rate}) requires some explanation. The flux of bound DM particles is proportional to  $g(r_m,e)T(r_m,e)\tau(r_m,e)$ as in the case of non-directional detectors divided by the effective area of the detector $\delta \ell_c^2$. Eventually we will show that the result will be independent of $\delta \ell_c$. In the case of non-directional detectors we were interested in the total flux of particles passing through the detector without caring about the direction. Therefore once we knew the density of bound particles per orbit, we had to integrate over all possible orbits (i.e. $r_m$ and $e$) in order to estimate the total rate. In the case of directional detection, not only do we care about the total number of events per time, but we need to know from what direction DM particles come from. Since we care about detecting particles that scatter off nuclei in the detector creating a nuclear recoil to a particular direction, $r_m$ and $e$ are not the only variables we need to achieve that. In addition to the characteristics of the elliptical orbit, we need to know what is the location of the perihelion of the orbit compared to the detector location.Therefore we parametrize the orbits by $r_m$,  $e$, the polar angles $\theta$ and $\phi$ that define the location of the perihelion with respect to the detector (i.e. the detector is along the $z$-axis) and the angle $\omega$ between the plane of the orbit and the plane defined by the perihelion the center of the Earth and the detector. We expect an isotropic distribution of the perihelion around the Earth and a uniform distribution for $\omega$. This is why we divide the corresponding quantites by $4 \pi$ and $2 \pi$ respectively in Eq.~(\ref{dir_rate}). The $\delta$ function enforces the recoil angle $\theta_q$ to be the one that kinematics dictates. We now need  to find the orbits that pass from the detector's location and can create a nuclear recoil to a particular horizontal direction. Eq.~(\ref{elip}) evaluated at $\theta=0$ gives the perihelion $r=r_m$. Therefore trading $P$ for $r_m$ and using $r=\ell_c$ (the distance of the detector from the center of the Earth) we rewrite Eq.~(\ref{elip}) as
\beq
\label{rmm}
 r_m=\ell_c\frac{1+e \cos\theta}{1+e}.
\eeq
For a given orbit where the perihelion forms an angle $\theta$ with the center of the Earth and the detector, $r_m$ must be given by the above equation in order for the particle to pass from the detector's location. Varying the value of the perihelion while keeping $e$ and $\theta$ fixed leads to
\beq
\label{drm}
\delta r_m=\delta \ell_c\frac{1+e \cos\theta}{1+e}.
\eeq
The integration over $dr_m$ can be substituted approximately by $\delta r_m$ which is related to the size of the detector. On the other hand in order for the orbit to pass through the detector (of dimension $\delta\ell_c$), 
\beq
\label{dw}
\ell_c \sin\theta \delta\omega=\delta\ell_c \Rightarrow \delta\omega=\frac{\delta\ell_c}{\ell_c \sin\theta}.
\eeq
Since $\theta$ takes values from 0 to $\pi$, it is always positive. Since $\delta\ell_c<<\ell_c$ $\delta\omega$ is extremely small unless one considers very small values of $\theta$ (practically locating the perihelion inside the detector). If we ignore this tiny patch of surface for the perihelion, we can substitute the integration over $d\omega$ by $\delta\omega$. Using Eqs~(\ref{drm}) and (\ref{dw}) we can write (\ref{dir_rate}) as
\begin{widetext}
\beq
\label{dir_rate2}
\frac{dR}{dE_Rd\Omega_q}=\frac{N_T}{16 \pi^3  \ell_c} \int \frac{d\sigma_N}{dE_R}g(r_m,e)\frac{T(r_m,e)}{\tau(r_m,e)}\delta \left ( \cos\theta_q -\frac{v_{\text{min}}}{v} \right )\frac{1+e z}{1+e}\frac{1}{\sqrt{1-z^2}}dzd\phi de,
\eeq
\end{widetext}
where $r_m$ is given by Eq.~(\ref{rmm}). We defined $z\equiv \cos\theta$. Note that the rate does not depend anymore on the characteristic size of the detector $\delta\ell_c$. We will eventually use the delta function to perform the integral over $z$. Before we do this, we need to find the relation of $\theta_q$ with the variables of the problem i.e. $e$, $\phi$ and $z$. Let us consider for the moment an orbit with $\phi=0$ and an angle $\theta$ subtended by the detector, the center of the Earth and the perihelion of the orbit. If we use cartesian coordinates with the perihelion being along the $x$-axis,  a point in the orbit has  coordinates 
\beq 
\label{xxx}
x=a_0 e +r\cos\theta,~~ ~~ y=r\sin\theta
\eeq with $a_0$ being the focal point. Let us choose a horizontal direction at the location of the detector
\beq
\label{hattheta}
\hat{\theta}=-\sin\theta \hat{x}+\cos\theta \hat{y}.
\eeq
A bound DM particle that follows a particular elliptical orbit reaches the detector with a velocity that has a direction
\beq
\label{hatel}
\hat{\ell}=\frac{dx}{d\ell}\hat{x}+\frac{dy}{d\ell}\hat{y}.
\eeq
With the help of Eq.~(\ref{xxx})
\beq
\frac{dx}{d\ell}=\frac{dr \cos\theta -r\sin\theta d\theta}{\sqrt{dx^2+dy^2}}=\frac{d\theta \left (\frac{dr}{d\theta} \cos\theta -r\sin\theta \right )}{d\theta\sqrt{\left (\frac{dr}{d\theta}\right )^2+r^2}}.
\eeq
Canceling the $d\theta$ from numerator and denominator and calculating $dr/d\theta$ from Eq.~(\ref{elip}) we get the final result
\beq
\label{dxl}
\frac{dx}{d\ell}=-\frac{\sin\theta}{\sqrt{1+e^2+2e\cos\theta}}.
\eeq
Similarly 
\beq
\label{dyl}
\frac{dy}{d\ell}=\frac{e+\cos\theta}{\sqrt{1+e^2+2e\cos\theta}}.
\eeq
Using Eqs.~(\ref{hattheta}), (\ref{hatel}), (\ref{dxl}) and (\ref{dyl}) we get
\beq
\cos\theta_q=\hat{\theta}\cdot\hat{\ell}=\pm \frac{1+e \cos\theta}{\sqrt{1+e^2+2e\cos\theta}}.
\eeq
Recall that $\hat{\theta}$ is the recoil direction and $\hat{\ell}$ the direction of the velocity of the bound DM particle. The $\pm$ refers to the two possibilities
that the particle is orbiting the ellipse (counter)clockwise. It is not difficult to show that for a nonzero value of $\phi$ 
\beq
\label{thetaq2}
\cos\theta_q=\hat{\theta}\cdot \hat{\ell}=\pm \frac{1+e \cos\theta}{\sqrt{1+e^2+2e\cos\theta}}\cos\phi.
\eeq
We show the details of the derivation in the case of nonzero $\phi$ in the appendix. We assume that there is equally probable to have clockwise or counterclockwise orbits. Let us consider first the orbits with a plus sign in Eq.~(\ref{thetaq2}). We will multiply the corresponding rate by a factor of 1/2 since there is $50\%$ probability. In order to evaluate the $dz$ integration using the delta function in Eq.~(\ref{dir_rate2}), we will use the well known property
\beq
\delta [h(z)]=\frac{\delta (z-z_0)}{|h'(z_0)|},
\eeq
where $h(z)$ is a function of $z$, $z_0$ is the solution of the equation $h(z)=0$ and $h'(z_0)$ is the derivatize of $h(z)$ with respect to $z$ evaluated at $z_0$.
In our particular case 
\begin{align}
\label{hh}
h(z)&=\cos\theta_q -\frac{v_{\text{min}}}{v} \nonumber \\
&=\frac{(1+e z)\cos\phi}{\sqrt{1+e^2+2 e z}}-\frac{v_{\text{min}}}{\sqrt{\frac{2GM_{\oplus}}{\ell_c}}\sqrt{1-\frac{1-e^2}{2(1+e z)}}},
\end{align}
where $v$ is given by Eq.~(\ref{vv}) with $r_m$ given by Eq.~(\ref{rmm}). Recall that $z=\cos\theta$. The equation $h(z)=0$ has the solution
\beq
\label{z0}
z_0=\frac{-\cos^2\phi+\gamma}{e\cos^2\phi},
\eeq
where
\beq
\label{gamma}
\gamma=\frac{v_{\text{min}}^2\ell_c}{GM_{\oplus}}.
\eeq
It is also easy to show that
\beq
|h'(z_0)|=\frac{e\cos^2\phi}{2\sqrt{2\gamma -(1-e^2)\cos^2\phi}}.
\eeq
The constraint $-1<z_0<1$ leads to the condition
\beq
\label{phiphi}
\sqrt{\frac{\gamma}{1+e}}<\cos\phi<\sqrt{\frac{\gamma}{1-e}}.
\eeq
From Eq.~(\ref{hh}) it is clear that $0<\cos\phi<1$ and therefore $\sqrt{\gamma/(1+e)}<1$. This last condition can be rewritten as
\beq
\label{e1}
e>\gamma-1.
\eeq
Recall that $0<e<1$ and therefore to have a nonzero signal $\gamma-1<1 \Rightarrow \gamma<2$. Using the definition of $\gamma$ (Eq.~(\ref{gamma})) and $v_{\text{min}}$ from Eq.~(\ref{vmin}), the constraint $\gamma < 2$ becomes
\beq
\label{ERc}
E_R<\frac{4 \mu_N^2GM_{\oplus}}{m_N \ell_c}.
\eeq
This condition in fact sets the upper limit in the recoil energy spectrum that bound DM particles can contribute.

\begin{figure*}[t]
\begin{minipage}{0.49\textwidth}
\begin{center}
\includegraphics[width=\textwidth]{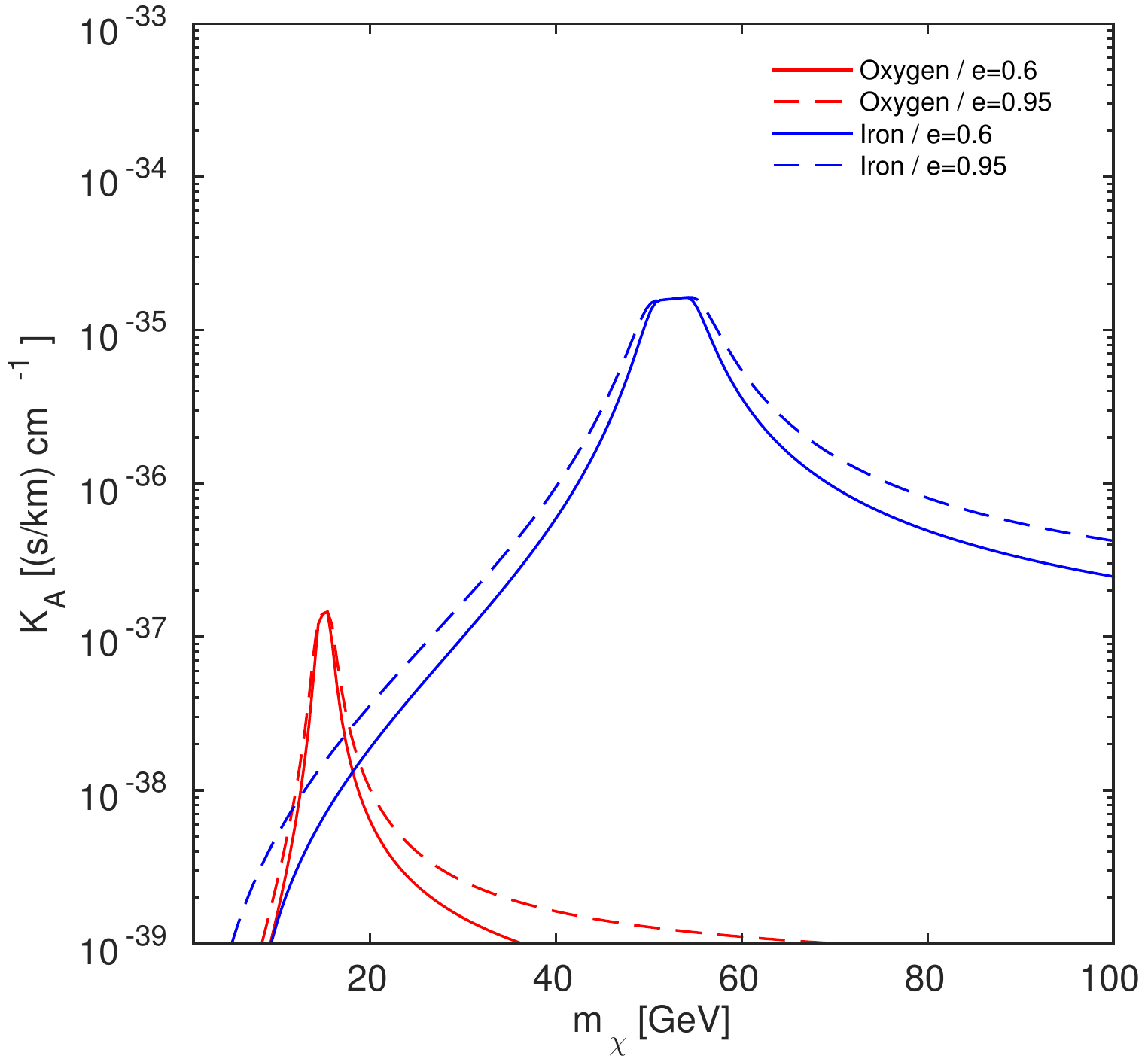}
\end{center}
\end{minipage}
\begin{minipage}{0.49\textwidth}
\begin{center}
\includegraphics[width=\textwidth]{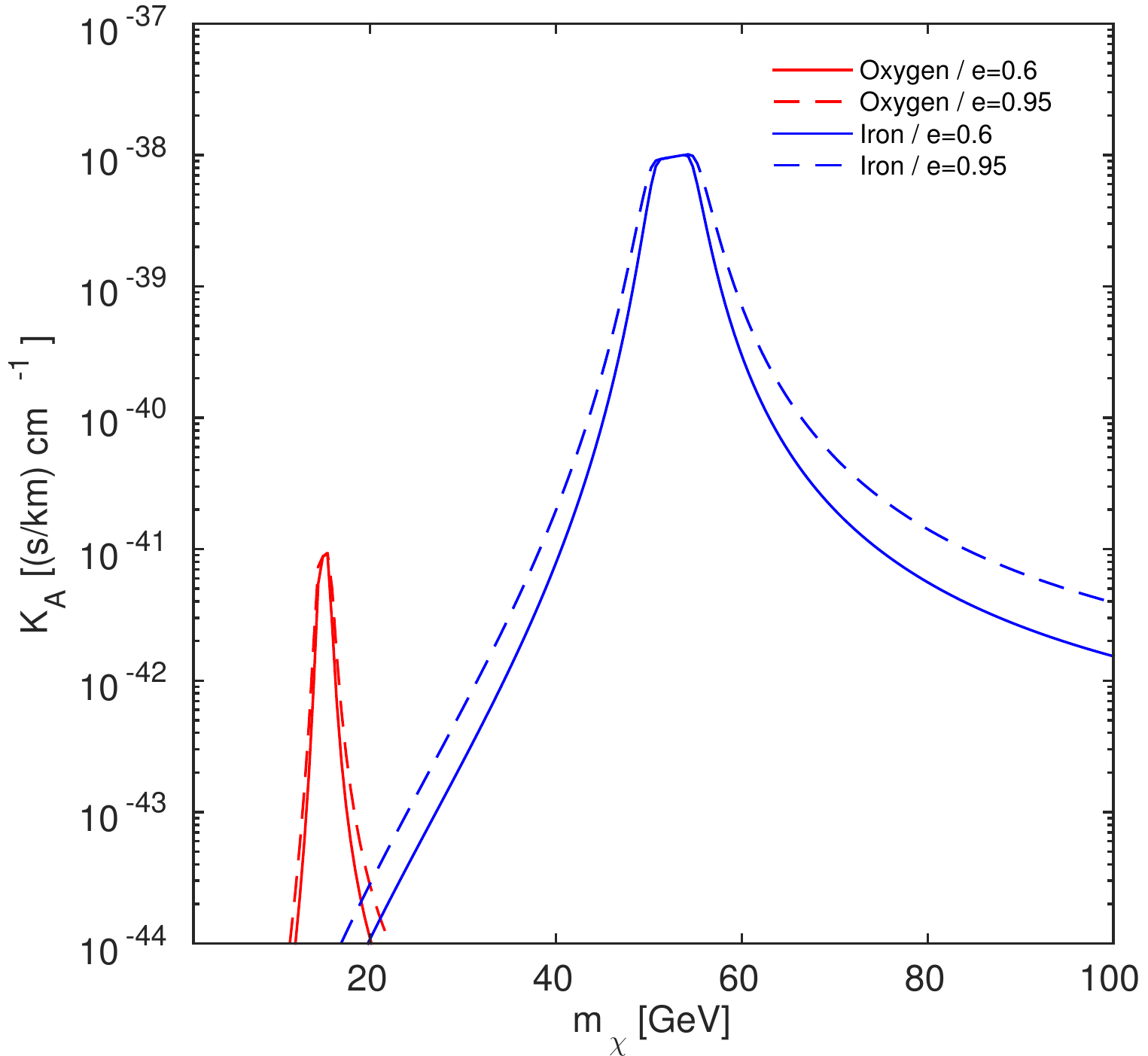}
\end{center}
\end{minipage}
\caption{$K_A$ as a function of the DM particle mass $m_\chi$ for two elements in the Earth, namely Oxygen and Iron.~$K_A$ is proportional to the probability of scattering towards a bound orbit of given ellipticity $e$ and perihelion $r_m$.~In the figure we vary $e$ as reported in the legends, and fix $r_m$ to $R_{\oplus}/2$.~The left panel refers to the interaction $\mathcal{O}_1$ with $c_1^0=2/m_V^2$ and $c_1^1=0$, whereas the right panel to the interaction $\mathcal{O}_{11}$ with $c_{11}^0=2/m_V^2$ and $c_{11}^1=0$.~The parameter $m_V=246.2$~GeV corresponds to the electroweak scale.}
\label{fig:K}
\end{figure*}

We can now rewrite Eq.~(\ref{dir_rate2}) performing the integration over $z$ by using the delta function as we prescribed above
\begin{widetext}
\beq
\label{dir_rate3}
\frac{dR}{dE_Rd\Omega_q}=\frac{N_T}{4 \pi^3  \ell_c}\int_{e_1}^1 \int_{\phi_1}^{\phi_2} \frac{d\sigma_N}{dE_R}g(r_m,e)\frac{T(r_m,e)}{\tau(r_m,e)}\frac{1+e z_0}{1+e}\frac{1}{\sqrt{1-z_0^2}}\frac{\sqrt{2\gamma-(1-e^2)\cos^2\phi}}{\cos^2\phi}\Theta(2-\gamma)d\phi de.
\eeq
\end{widetext}
$e_1=\text{Max}[\gamma-1,0]$ is derived from the constraint of Eq.~(\ref{e1}) and the fact that $e>0$. The step function $\Theta(2-\gamma)$ ensures that $\gamma<2$ as it is required from the constraint of $(\ref{ERc})$. The constraint of Eq.~(\ref{phiphi}) determines the limits of integration for $\phi$
\begin{align}
\phi_1&=\cos^{-1} \left [\text{Min} \left (1,\sqrt{\frac{\gamma}{1-e}} \right )\right ] \nonumber \\
\phi_2&=\cos^{-1}\sqrt{\frac{\gamma}{1+e}}.
\end{align}
Note that $r_m$ is evaluated at the value
\beq
r_m=\ell_c\frac{1+e z_0}{1+e},
\eeq
(see Eq.~(\ref{rmm})).
Eq.~(\ref{dir_rate3}) is our final result for the recoil spectrum in directional detectors. Comparing the overall coefficient of Eq.~(\ref{dir_rate3}) with respect to that of Eq.~(\ref{dir_rate2}), the former is larger by a factor of 4 (there is a factor of $1/4$ versus $1/16$ respectively). A factor of 2 comes from the integration of $\phi$. Note that Eq.~(\ref{phiphi}) is satisfied in two regions i.e. one with positive and one with negative value of $\phi$. Since $\cos\phi$ always appears as a square, we integrate only over positive $\phi$ and multiply by 2. The second factor of 2 comes from the fact that the orbits with the opposite direction (i.e. with a minus sign in Eq.~(\ref{thetaq2})) give exactly the same contribution as the orbits with the plus sign. This is easy to show: The solution of Eq.~(\ref{hh}) is still given by (\ref{z0}) even for the orbits with a minus sign in (\ref{thetaq2}). The only difference is that in this case $\cos\phi<0$. The constraint of Eq.~(\ref{phiphi}) remains the same once $\cos\phi\rightarrow -\cos\phi$. However since $\cos\phi$ appears always as $\cos^2\phi$ in Eq.~(\ref{dir_rate3}), one can change variable $\phi'\equiv \pi-\phi$ keeping in mind that $\cos^2\phi'=\cos^2\phi$. The constraint on $\phi'$ is the same of Eq.~(\ref{phiphi}) with $\phi\rightarrow \phi'$ since $\cos\phi'=-\cos\phi$. The value of $|h'(z_0)|$ is the same as before and therefore the overall contribution of the ``negative sign" orbits is the same as the ones with positive sign. 

\begin{figure*}[t]
\begin{minipage}{0.49\textwidth}
\begin{center}
\includegraphics[width=\textwidth]{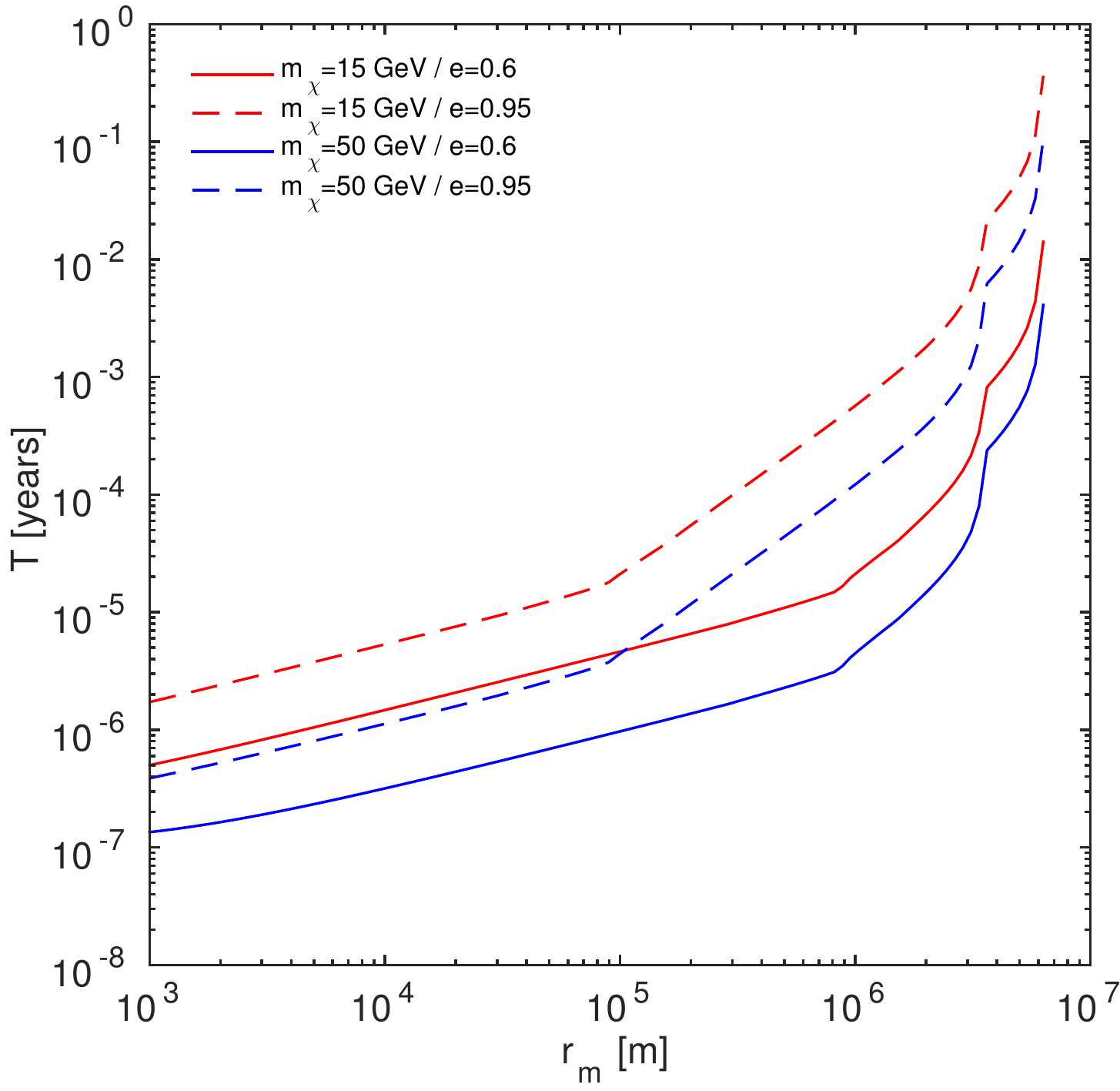}
\end{center}
\end{minipage}
\begin{minipage}{0.49\textwidth}
\begin{center}
\includegraphics[width=\textwidth]{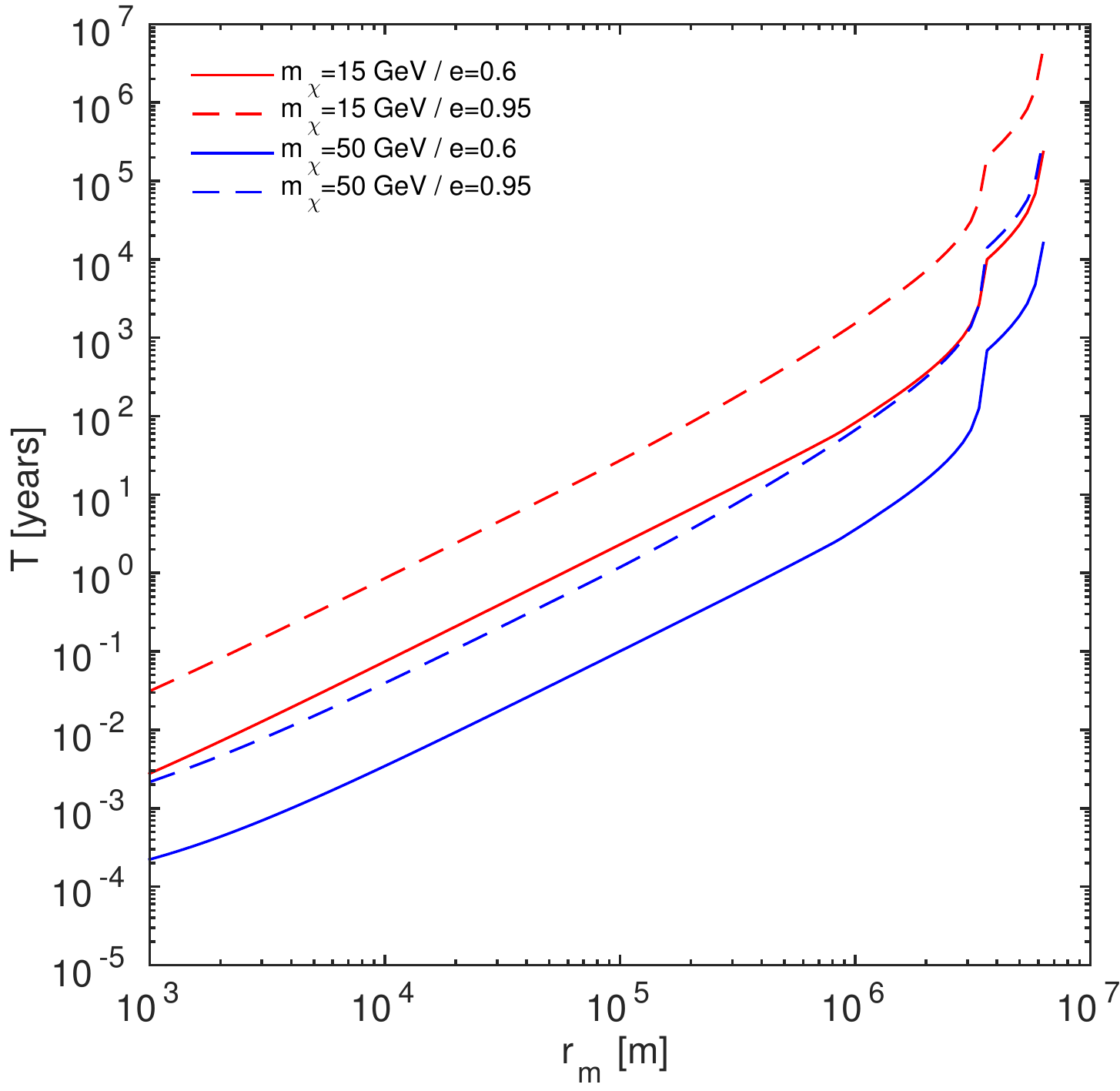}
\end{center}
\end{minipage}
\caption{Accumulation time $T$ as a function of the perihelion $r_m$ for two reference values of the DM particle mass $m_\chi$ and of the ellipticity $e$.~The left panel refers to the interaction $\mathcal{O}_1$ with $c_1^0=2/m_V^2$ and $c_1^1=0$, whereas the right panel to the interaction $\mathcal{O}_{11}$ with $c_{11}^0=2/m_V^2$ and $c_{11}^1=0$.~Coupling constants are expressed in terms of the electroweak scale $m_V=246.2$~GeV.~The accumulation time for the operator $\mathcal{O}_{11}$ is significantly larger than that of $\mathcal{O}_{1}$.~Overall, $T\times K_A$ for $O_{11}$ is larger than the corresponding of $O_1$.}
\label{fig:T}
\end{figure*}

Eq.~(\ref{dir_rate3}) can be used for any generic form of DM-nucleon interactions. For the  spin-independent interaction of Eq.~(\ref{SIcs}), (\ref{dir_rate3}) becomes 
 \begin{align}
\label{direct2}
\frac{dR}{dE_R d \Omega_q}= &\kappa_d \int_{e_1}^1 de \int_{\phi_a}^{\phi_b} d\phi \frac{1}{v^2}\frac{g(r_m,e)}{\tau(r_m,e)} T(r_m,e)  \nonumber \\ &\times \frac{1+e z_0}{1+e} \frac{1}{\sqrt{1-z_0^2}}
 \frac{\sqrt{2\gamma-(1-e^2)\cos^2\phi}}{e \cos^2\phi},
\end{align}
where $\kappa_d=N_T m_N \sigma_n A_N^2F^2(E_R)/(8 \pi^3 \mu_N^2 \ell_c)$.

Eq.~(\ref{dir_rate3}) describes the recoil spectrum of bound DM scattering off an underground detector. This spectrum must be contrasted to the usual directional spectrum of halo DM. Using
Eq.~(\ref{dir_cross}) we get
\beq
\label{eq:halodir}
\frac{dR}{dE_Rd \Omega_q}=\frac{N_T n_{\chi}}{2 \pi} \int \frac{d\sigma_N}{dE_R} \delta \left ( \hat{v} \cdot \hat{q}-\frac{v_{\text{min}}}{v} \right )f(v)v d^3v.
\eeq In the case of spin indeppendent interactions (see Eq.~(\ref{SIcs})), it takes the form
\beq
\frac{dR}{dE_Rd \Omega_q}=\kappa_h \hat{f}(v_{\text{min}},\hat{q}),
\eeq
where $\kappa_h=N_Tn_{\chi}m_N\sigma_n A_N^2 F_N^2(E_R)/(4 \pi \mu_N^2)$ and $\hat{f}(v_{\text{min}},\hat{q})$ is the so-called Radon transfromation of $f(v)$ defined as~\cite{Gondolo:2002np}
\beq
\hat{f}(v_{\text{min}},\hat{q})=\int \delta (\vec{v}\cdot \hat{q}-v_{\text{min}})f(v)d^3v.
\eeq

\section{Results}
\label{sec:results}
The main equations derived in the previous sections are Eqs.~(\ref{eq:rec}) and (\ref{dir_rate3}).~They describe the rate of nuclear recoil events expected in non-directional and directional detectors, respectively.~Now we numerically evaluate and interpret these expressions under different assumptions regarding the cross-sections $d\sigma_A/dE_R$ (for scattering in the Earth) and $d\sigma_N/dE_R$ (for scattering in a terrestrial detector).~We will also investigate the dependence of our results on the type of target nuclei composing the detector in analysis.\\

\subsection{General considerations}
\label{sec:conclusion}
The rate of nuclear recoil events in Eq.~(\ref{eq:rec}) depends on the cross-section $d\sigma_A/dE_R$ through the functions $K_A$ and $T$ (defined in Eqs.~(\ref{eq:K}) and (\ref{eq:Gamma}), respectively).~It also depends on the differential cross-section $d\sigma_N/dE_R$, which appears in Eq.~(\ref{eq:rec}) directly.~We can therefore characterise each single scattering event at detector as the result of a complex three stage physical process.~Each stage explicitly depends on how DM interacts with nuclei and is briefly described below:
\begin{enumerate}
\item Capture of the DM particle $\chi$ by the Earth.~The element $A$ contributes with probability proportional to $K_A$.
\item Motion of the DM particle $\chi$ along the bound orbit characterised by $r_m$ and $e$.~This motion lasts on average for a time $T$, i.e.~the accumulation time defined in Eq.~(\ref{eq:Gamma}).
\item Scattering of the particle $\chi$ at detector (with cross-section given by $d\sigma_N/dE_R$).
\end{enumerate}
In all numerical applications, we will assume the cross-section
\begin{align}
\frac{d \sigma_A}{d E_R}
\label{eq:sigma}  & = \frac{2 m_A}{(2 j_A +1) v^2} \sum_{\tau=0,1} \sum_{\tau'=0,1}  \Big[ c_1^\tau c_1^{\tau'} W_{M}^{\tau\tau'}(E_R) 
\nonumber \\ 
& + \frac{2 j_\chi(j_\chi + 1)}{3} \, \frac{m_A E_R}{m_n^2} \, c_{11}^\tau c_{11}^{\tau'} W_{\Phi''}^{\tau\tau'}(E_R) \Big] \,,
\end{align}
\begin{figure*}[t]
\begin{minipage}{0.49\textwidth}
\begin{center}
\includegraphics[width=\textwidth]{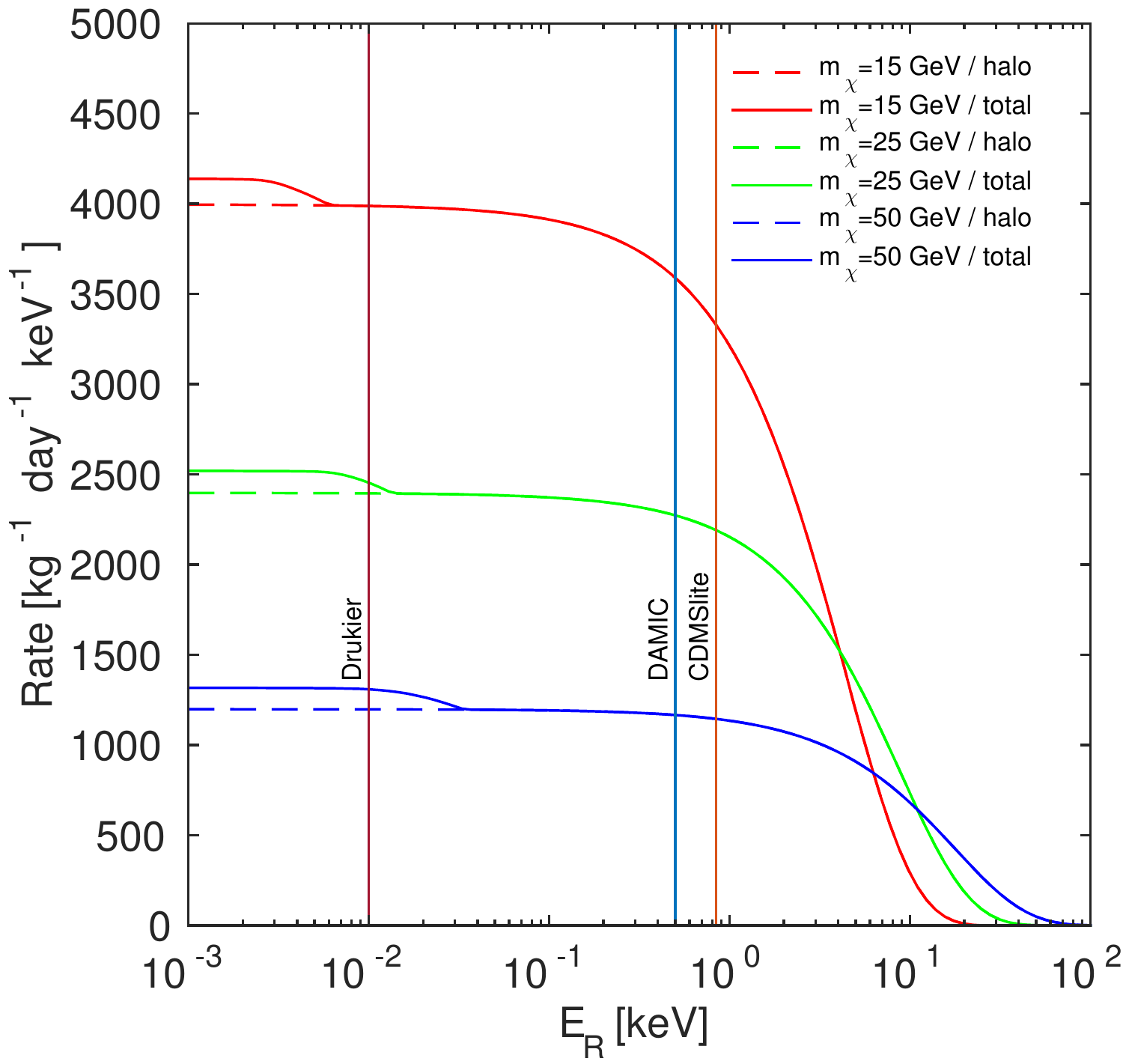}
\end{center}
\end{minipage}
\begin{minipage}{0.49\textwidth}
\begin{center}
\includegraphics[width=\textwidth]{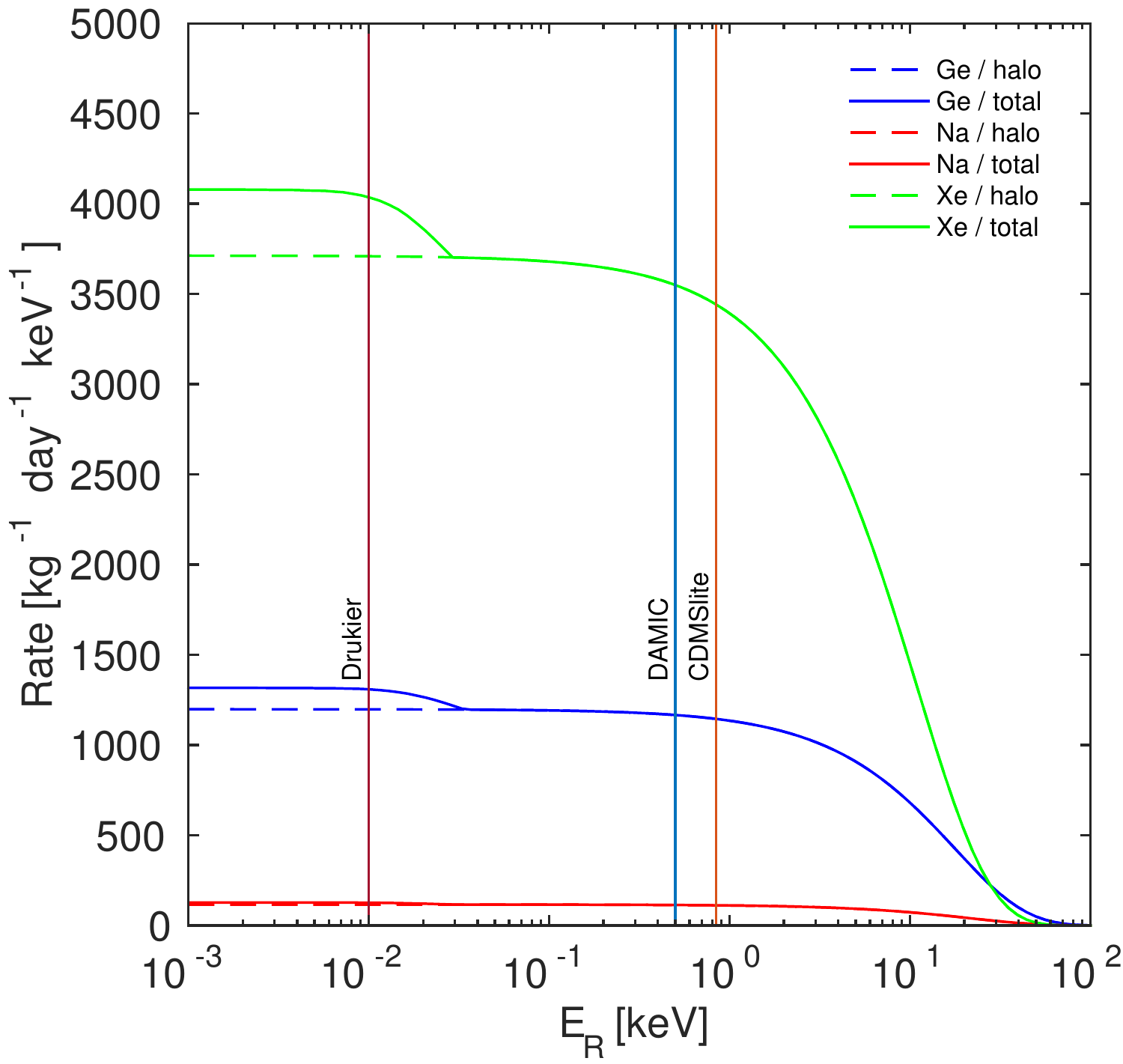}
\end{center}
\end{minipage}
\caption{Rate of nuclear recoil events $dR/dE_R$ as a function of $E_R$.~We assume dark matter-nucleon interactions of type $\mathcal{O}_1$ and $c_1^0=2/m_V^2$, $c_1^1=c_{11}^0=c_{11}^1=0$ ($m_V=246.2$~GeV).~The left panel reports results obtained for three different values of the DM particle mass, and assuming Germanium as a target material.~In the right panel we fix $m_\chi=50$~GeV, and consider different target materials for $d\sigma_N/dE_R$, namely, Xenon, Germanium and Sodium.~In both panels, solid lines correspond to the total rates, including the contribution from halo and bound DM particles.~Dashed lines represent the contribution to $dR/dE_R$ from halo DM particles.~Vertical lines show illustrative energy thresholds of running or proposed dark matter direct detection experiments.}
\label{fig:rate_op1}
\end{figure*}and an analogous expression for $d\sigma_N/dE_R$.~The isotope-dependent nuclear response functions $W_M^{\tau\tau'}$ and $W_{\Phi''}^{\tau\tau'}$ in Eq.~(\ref{eq:sigma}) are quadratic in nuclear matrix elements and are defined in Ref.~\cite{Fitzpatrick:2012ix}.~They have been calculated for the 16 most abundant elements in the Sun, including $^{16}$O, $^{28}$Si, $^{24}$Mg, $^{56}$Fe, $^{40}$Ca, $^{23}$Na, $^{32}$S, $^{59}$Ni, and $^{27}$Al, in Ref.~\cite{Catena:2015uha} and for various isotopes of Xe and Ge, and for Na in Ref.~\cite{Anand:2013yka}.~The labels $1$ and $11$ in Eq.~(\ref{eq:sigma}) refer to the non-relativistic effective operators $\mathcal{O}_{1}$ and $\mathcal{O}_{11}$ introduced in Ref.~\cite{Fitzpatrick:2012ix}.~The former corresponds to the familiar spin-independent interaction operator, the latter to the momentum-dependent interaction operator $\mathcal{O}_{11}=(\mathbf{q}/m_n) \cdot \mathbf{S_\chi}$, where $m_n$ is the nucleon mass, and $\mathbf{q}$ and $\mathbf{S_\chi}$ are the momentum transfer and DM particle spin operators, respectively.~They are explicitly defined in Ref.~\cite{Catena:2015uha}.~A comparison of Eqs.~(\ref{eq:sigma}) and (\ref{SIcs}) allows to express $\sigma_A$ and $F_A$ in terms of the coupling constants and response functions in Eq.~(\ref{eq:sigma}).~For the isoscalar coupling constants, $c_1^0$ and $c_{11}^0$, we assume the reference values $2/m_V^2$, with $m_V=246.2$~GeV (the electroweak scale), or $0$, depending on whether we are interested in the operator $\mathcal{O}_{1}$ or $\mathcal{O}_{11}$.~At the same time, we set the isovector coupling constants to zero:~$c_1^1=c_{11}^1=0$.~Finally, $j_A$ and $j_\chi$ are the $A$ element and DM particle spins, respectively.

\begin{figure*}[t]
\begin{minipage}{0.49\textwidth}
\begin{center}
\includegraphics[width=\textwidth]{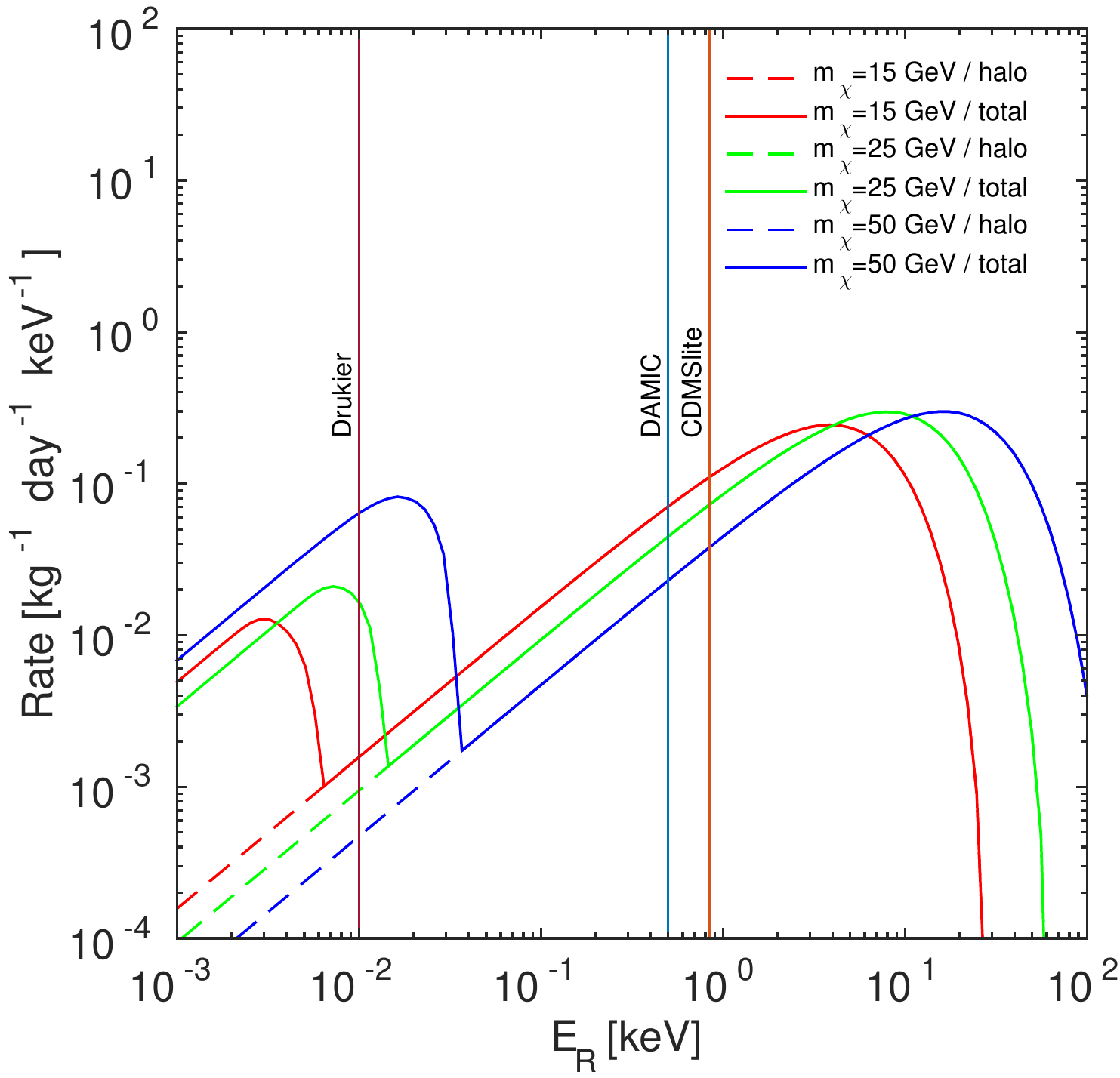}
\end{center}
\end{minipage}
\begin{minipage}{0.49\textwidth}
\begin{center}
\includegraphics[width=\textwidth]{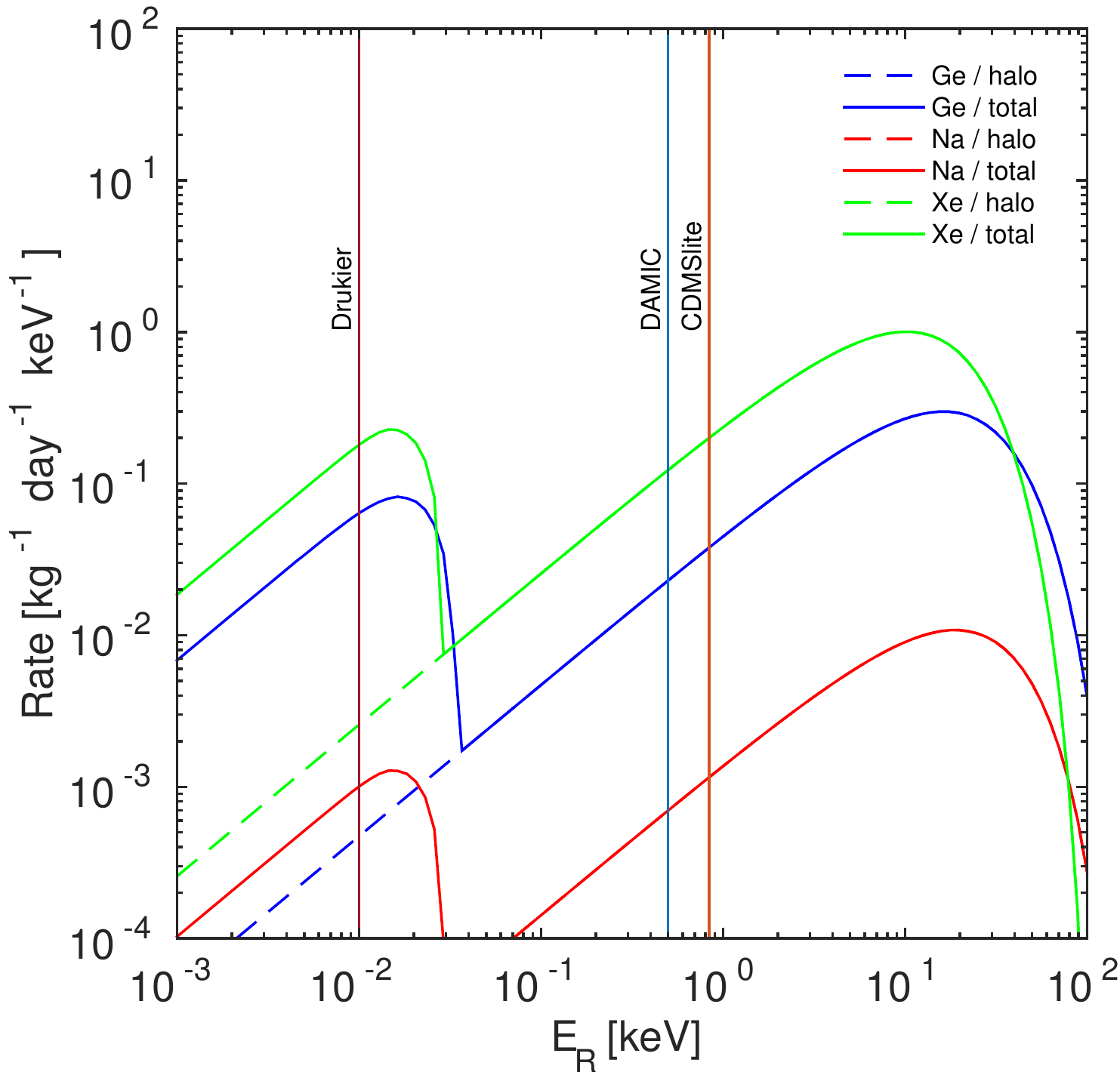}
\end{center}
\end{minipage}
\caption{Same as for Fig.~\ref{fig:rate_op1} but now for the interaction $\mathcal{O}_{11}$.}
\label{fig:rate_op11}
\end{figure*} 

\begin{figure*}[t]
\begin{minipage}{0.49\textwidth}
\begin{center}
\includegraphics[width=\textwidth]{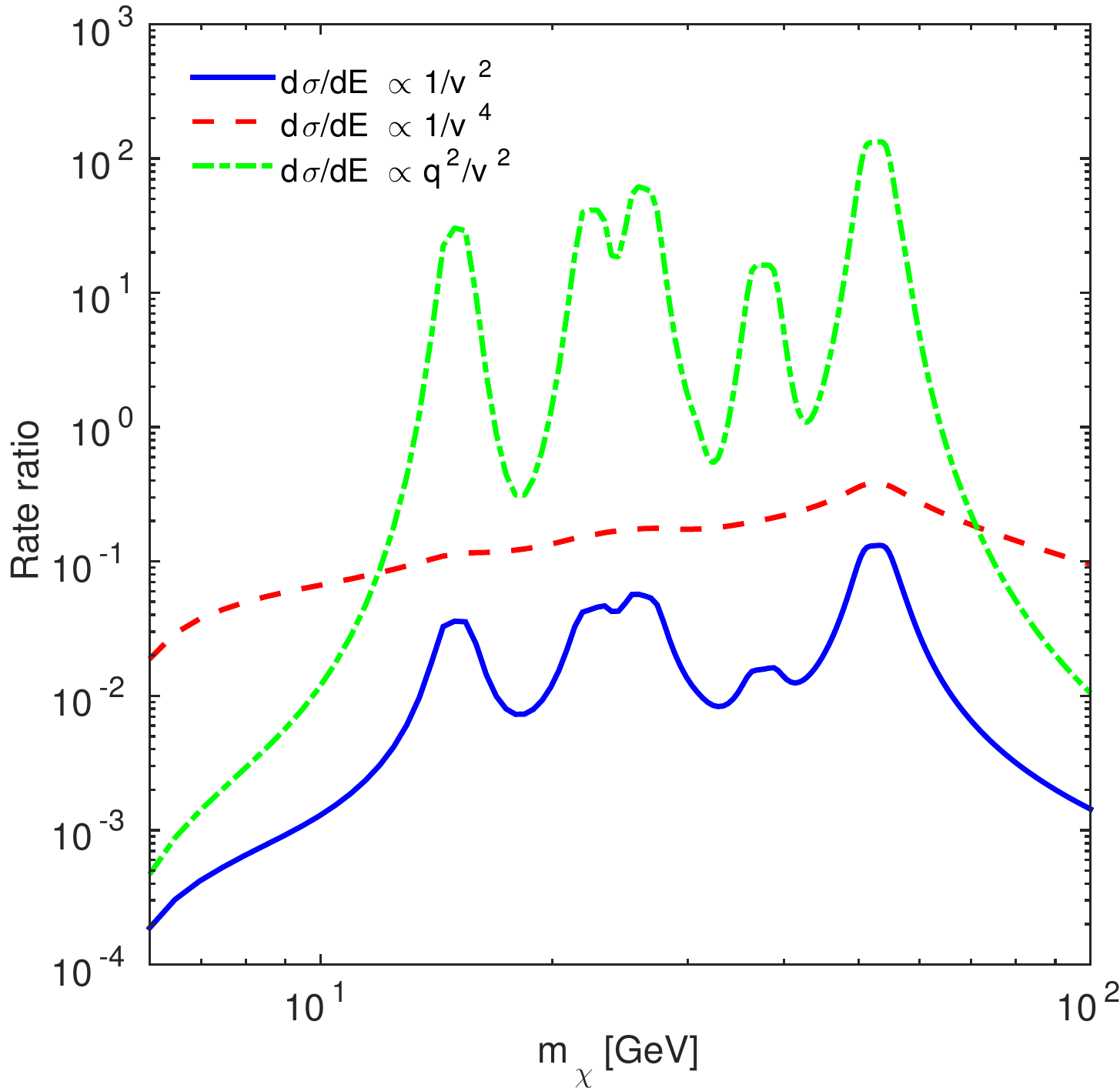}
\end{center}
\end{minipage}
\begin{minipage}{0.49\textwidth}
\begin{center}
\includegraphics[width=\textwidth]{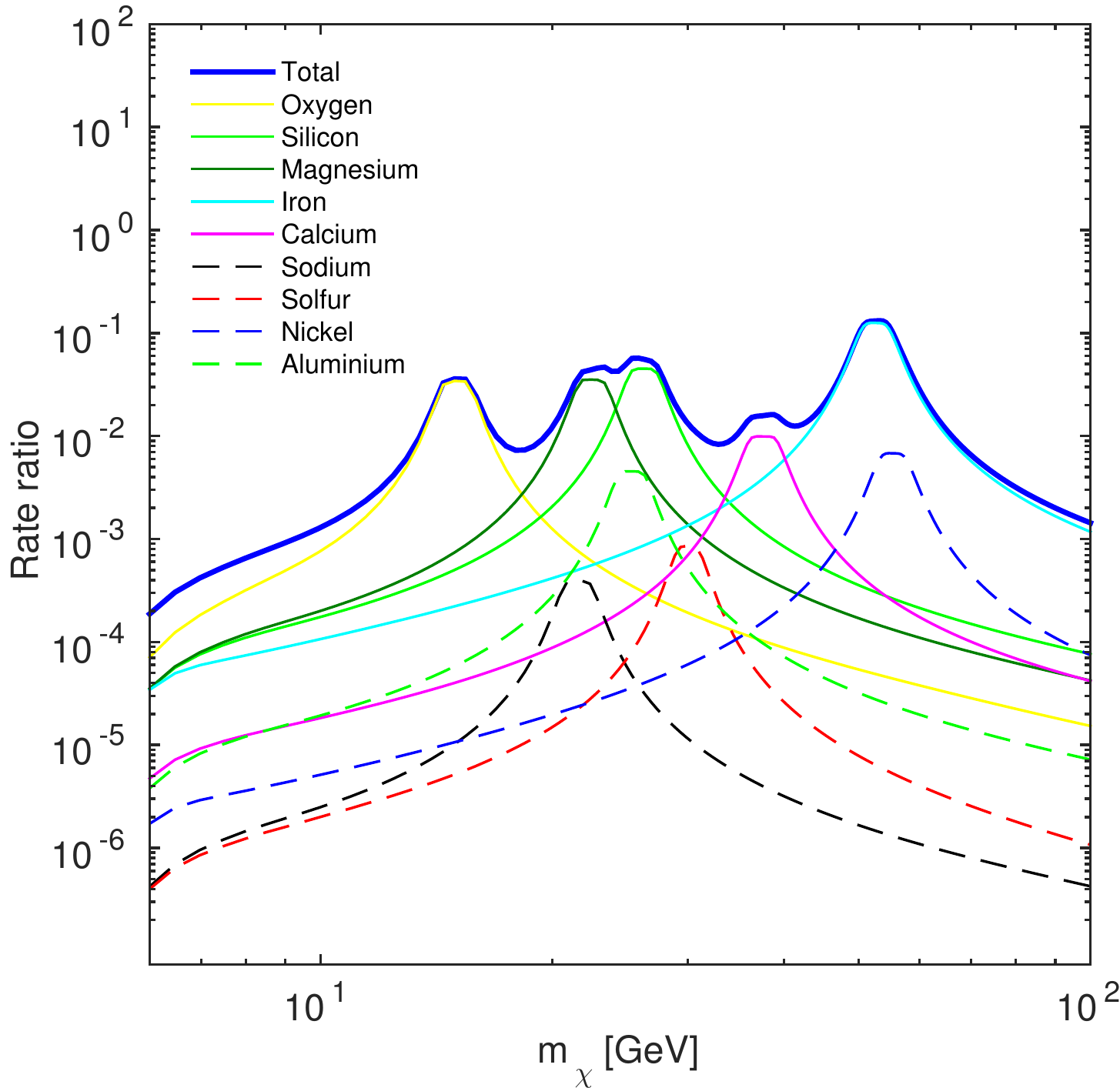}
\end{center}
\end{minipage}
\caption{Left:~Ratio of Eqs.~(\ref{eq:rec}) and (\ref{eq:halo}) as a function of $m_\chi$ for three different dark matter-nucleon interactions.~From the top to the bottom in the legend:~$\mathcal{O}_1$, a modified version of $\mathcal{O}_1$ obtained by replacing $c_1^0$ with $c_1^0/v$  (i.e.~resonant scattering), and  $\mathcal{O}_{11}$.~In all cases we set $c_{11}^1$ and $c_1^1$ to zero, and assume Germanium as a target material.~In the figure, we introduce the symbol $d\sigma/dE$ to characterise the scaling of $d\sigma_A/dE_R$ and $d\sigma_N/dE_R$ as a function of the dark matter-nucleus relative velocity $v$ and of the momentum transferred $q$.~Right:~Contribution of $^{16}$O, $^{28}$Si, $^{24}$Mg, $^{56}$Fe, $^{40}$Ca, $^{23}$Na, $^{32}$S, $^{59}$Ni, and $^{27}$Al to the ratio of Eqs.~(\ref{eq:rec}) and (\ref{eq:halo}) as a function of the DM mass $m_\chi$.~We assume $\mathcal{O}_1$ as dark matter-nucleon interaction, $c_1^0=2/m_V^2$ ($m_V=246.2$~GeV) and $c_1^1=0$.~In both panels we assume a nuclear recoil energy of 1 eV in the evaluation of the scattering rates.}
\label{fig:ratio}
\end{figure*}

Knowledge of the Earth's chemical composition is needed in order to evaluate $K_A$ and $T$.~In this study, we consider the nine elements:~$^{16}$O, $^{28}$Si, $^{24}$Mg, $^{56}$Fe, $^{40}$Ca, $^{23}$Na, $^{32}$S, $^{59}$Ni, and $^{27}$Al, with mass fractions as given in Ref.~\cite{massf}, and the radial density given in Ref.~\cite{dens} and implemented in Ref.~\cite{Gondolo:2004sc}.~We have verified numerically that changes in the mass fraction of single elements in the Earth have a negligible impact on the scattering rate evaluation.

Fig.~\ref{fig:K} shows $K_A$ as a function of the DM particle mass $m_\chi$ for two elements in the Earth, namely Oxygen and Iron, and for two reference values of the ellipticity $e$.~We find that $K_A$ increases for $m_\chi\rightarrow m_A$ since in this mass range the upper limit $v_2$ in Eq.~(\ref{eq:K}) tends to infinity,~i.e. maximum momentum transfer in the scattering.~Fig.~\ref{fig:K} also shows that for large values of $e$ the range of masses where $K_A\neq0$ is broader than for $e\simeq 0$.~The reason is that for a given $r_m$, the upper limit $v_2$ (Eq.~\ref{v2}) in the integral defining $K_A$ grows with $e$ and the integrand in Eq.~(\ref{eq:K}) is proportional to $v^3$ which also grows with $e$.~Fig.~\ref{fig:K} has been obtained by setting $r_m = R_\oplus/2$.

Fig.~\ref{fig:T} shows $T$ as a function of $r_m$ for two reference values of $m_\chi$ and $e$.~As expected, $T$ grows when $r_m \rightarrow~R_\oplus$ and $e\rightarrow 1$ since the intersection of these orbits with the Earth is small, which minimises the probability of a second DM scattering event.~In this work we assume that after a second scattering event, DM particles sink at the centre of the planet and cannot be detected directly.~Notice also that in all calculations discussed here, we assume elliptical orbits for the DM particles in bound orbits, which is rigorously correct only for trajectories external to the Earth.~It is however a fairly good approximation for orbits with $r_m \rightarrow~R_\oplus$ and $e\rightarrow 1$, i.e. for the orbits contributing the most to the rate of nuclear recoil events presented in what follows.~At the same time, Fig.~\ref{fig:T} is quantitatively reliable in the limit $r_m \rightarrow~R_\oplus$ and $e\rightarrow 1$ only.

\subsection{Non-directional detectors}
In this section we focus on the rate of nuclear recoil events $dR/dE_R$ in Eq.~(\ref{eq:rec}).~Fig.~\ref{fig:rate_op1} shows $dR/dE_R$ as a function of $E_R$ assuming dark matter-nucleon interactions of type $\mathcal{O}_1$.~We have obtained this figure under the assumption $c_1^0=2/m_V^2$, $c_{11}^0=0$.~The left panel reports results obtained for three different values of the DM particle mass and assuming Germanium as a target material, whereas in the right panel we consider different target materials for $d\sigma_N/dE_R$, namely Xenon, Germanium and Sodium, and fix the DM particle mass at $m_\chi=50$~GeV.~In both panels, solid lines are the total rates, including the contribution from halo and bound DM particles.~Dashed lines represent the contribution to $dR/dE_R$ from halo DM particles.~The case of dark matter-nucleon interactions of type $\mathcal{O}_{11}$ is discussed in Fig.~\ref{fig:rate_op11}, where in the left (right) panel we have reported results obtained for different DM particle masses (target materials).~In both figures, vertical lines correspond to the threshold energies of present (CDMSlite
\cite{Agnese:2013jaa} and DAMIC~\cite{Chavarria:2014ika}) or proposed (Drukier~\cite{Drukier:1983gj}) direct detection experiments.

From Figs.~\ref{fig:rate_op1} and \ref{fig:rate_op11} we conclude that DM particles in orbits bound to the Earth can be revealed in future direct detection experiments as pronounced features in the low-energy part of the induced nuclear recoil spectrum.~As in the case of halo DM, DM particles bound to the Earth can produce a larger number of nuclear recoil events at low-energies if they are light, and in detectors composed of heavy nuclei.

In order to assess the significance of the predicted spectral features, we evaluate the ratio of Eqs.~(\ref{eq:rec}) and (\ref{eq:halo}) as a function of $m_\chi$.~The result of this calculation is reported in Fig.~\ref{fig:ratio}.~The left panel shows the rate ratio for three dark matter-nucleon interaction types:~$\mathcal{O}_1$, a modified version of $\mathcal{O}_1$ obtained by replacing $c_1^0$ with $c_1^0/v$ in the equations above (e.g.~resonant scattering~\cite{Bai:2009cd}), and finally $\mathcal{O}_{11}$.

The rate ratio can be as large as 0.1 for the interaction $\mathcal{O}_1$, and 0.4 for its resonant analogous.~Notably, for the interaction $\mathcal{O}_{11}$ the value can be up to  $\sim$200 at the Iron resonance (i.e.~$m_\chi\sim50$~GeV).~The large value found for the operator $\mathcal{O}_{11}$ is related to the large accumulation time $T$ that DM particles interacting with nuclei via $\mathcal{O}_{11}$ can spend on bound orbits before a second scattering occurs (see right panel in Fig.~\ref{fig:T}).~We have verified numerically that the ratio of Eqs.~(\ref{eq:rec}) and (\ref{eq:halo}) is independent of the coupling constants $c_1^0$ and $c_{11}^0$ when a single interaction at the time is considered.

Finally, the right panel in Fig.~\ref{fig:ratio} shows the contribution of $^{16}$O, $^{28}$Si, $^{24}$Mg, $^{56}$Fe, $^{40}$Ca, $^{23}$Na, $^{32}$S, $^{59}$Ni, and $^{27}$Al to the ratio of Eqs.~(\ref{eq:rec}) and (\ref{eq:halo}) as a function of the DM particle mass, and assuming $\mathcal{O}_1$ as dark matter-nucleon interaction.~The overall shape of the rate ratio reflects the resonant form of the function $K_A$, and contributions from distinct elements in the Earth can easily be identified in the figure. 

\begin{figure*}[t]
\begin{minipage}{0.49\textwidth}
\begin{center}
\includegraphics[width=\textwidth]{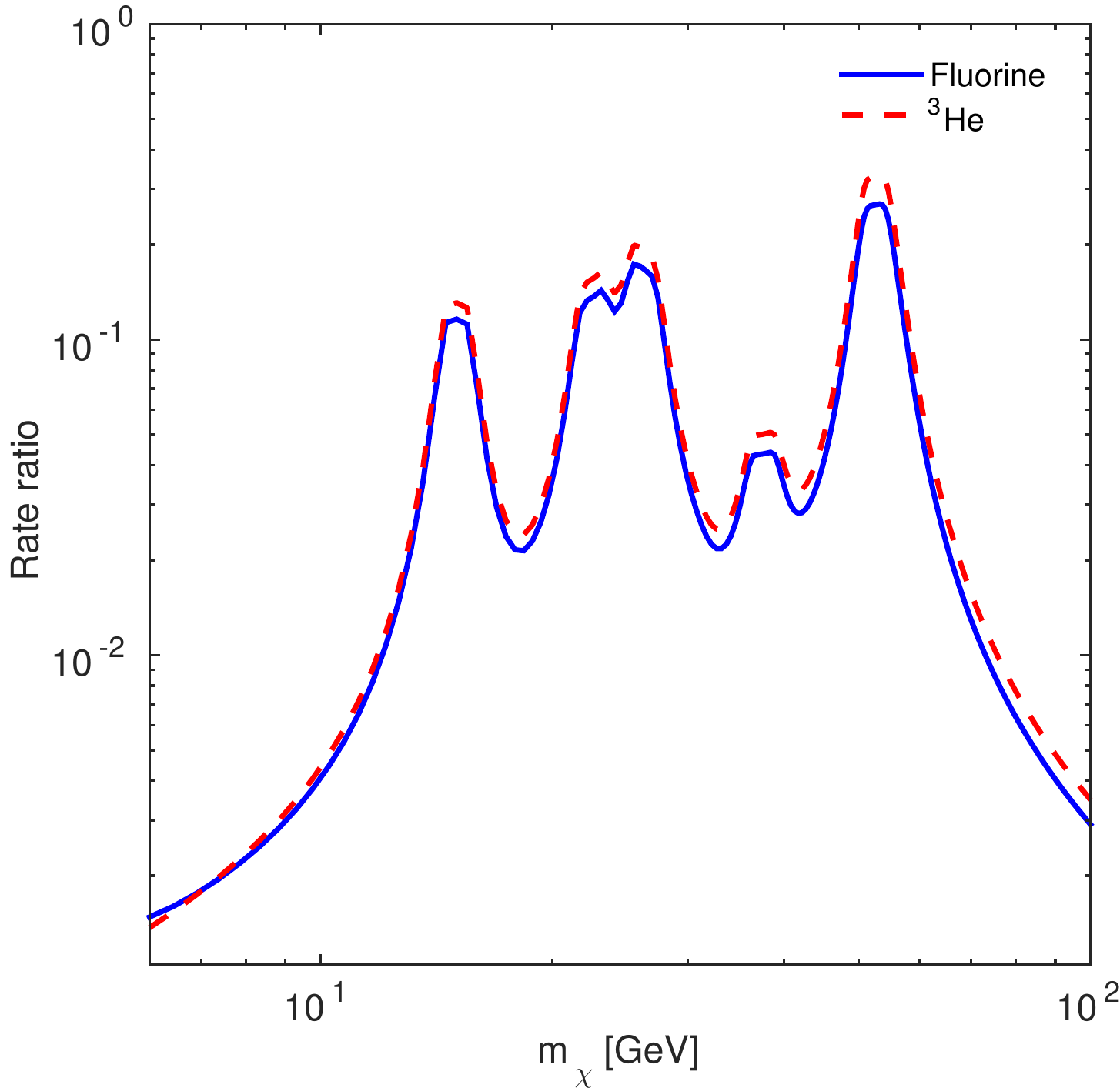}
\end{center}
\end{minipage}
\begin{minipage}{0.49\textwidth}
\begin{center}
\includegraphics[width=\textwidth]{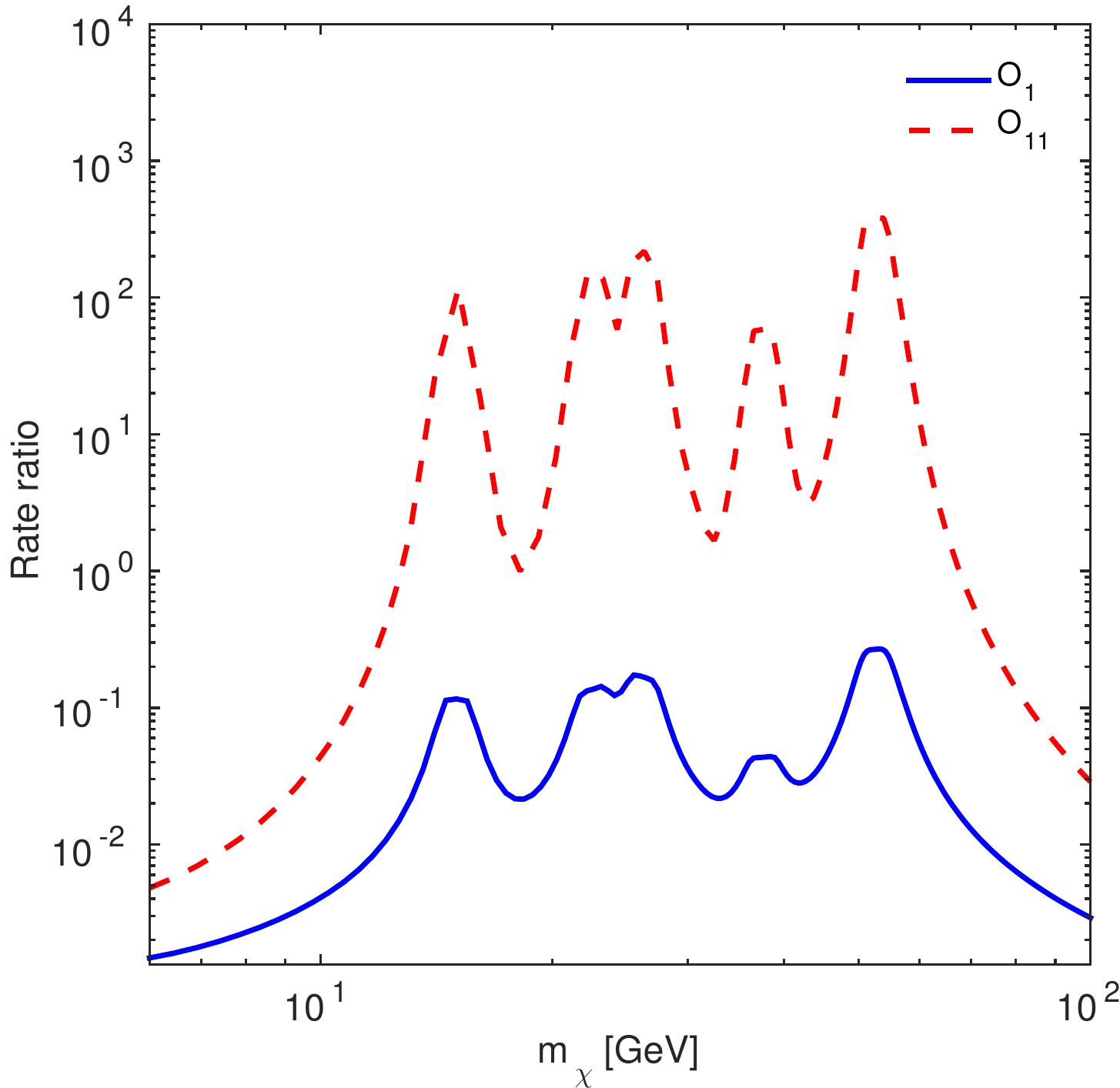}
\end{center}
\end{minipage}
\caption{Left:~Ratio of the double differential rates in Eqs.~(\ref{dir_rate3}) and (\ref{eq:halodir}) as a function of the DM particle mass.~We assume $\mathcal{O}_1$ as dark matter-nucleon interaction.~The solid blue line refers to a hypothetical detector composed of Fluorine, whereas the dashed red line corresponds to a second hypothetical detector which uses $^{3}$He as a target material.~Right:~Same ratio as in the left panel now evaluated for the operators $\mathcal{O}_1$ (blue solid line) and $\mathcal{O}_{11}$ (red dashed line) for comparison.~In both cases we assume Fluorine as a target material.~In the two panels we assume a nuclear recoil energy of 1 eV.}
\label{fig:dir}
\end{figure*}
\subsection{Directional detectors}
We conclude this section with a quantitative analysis of Eq.~(\ref{dir_rate3}), which describes the double differential rate of nuclear recoil events induced by the scattering of DM particles bound to the Earth in directional detection experiments~\cite{Catena:2015vpa,Kavanagh:2015jma,Kouvaris:2015laa}.

In Fig.~\ref{fig:dir}, the left panel shows the ratio of the double differential rates in Eqs.~(\ref{dir_rate3}) and (\ref{eq:halodir}) as a function of the DM particle mass, assuming $\mathcal{O}_1$ as dark matter-nucleon interaction.~The solid blue line refers to a hypothetical detector composed of Fluorine, whereas the dashed red line corresponds to a second hypothetical detector which uses $^{3}$He as a target material.~The right panel in Fig.~\ref{fig:dir} shows the same ratio, now evaluated for the operators $\mathcal{O}_1$ and $\mathcal{O}_{11}$, and assuming Fluorine as a target material.~As for the case of non-directional detectors, the predicted spectral feature is more pronounced for the operator $\mathcal{O}_{11}$ than for $\mathcal{O}_1$ by roughly three orders of magnitude.

For directional detectors the size of the effect is of the order of 0.1 ($10^2$) for $\mathcal{O}_1$ ($\mathcal{O}_{11}$), and it is generically larger than that of the non-directional detectors (e.g.~compare Fig.~\ref{fig:ratio} with Fig.~\ref{fig:dir}).~Furthermore, Fig.~\ref{fig:dir} shows that the predicted spectral feature is slightly more pronounced for a $^3$He based detector than for a detector adopting F as a target material.

\section{Conclusion}
\label{sec:conclusion}
We have studied the properties and detection prospects  of DM particles bound to the Earth.~The new DM population forms via scattering of Milky Way DM particles by nuclei in our planet's interior.~We have derived fluxes and nuclear recoil event rates at directional and non-directional detectors expected for the new population of DM particles.~The equations presented in this work are valid for arbitrary dark matter-nucleon interactions, and extend those found in Ref.~\cite{Catena:2016sfr}.~We have numerically evaluated such expressions under different assumptions regarding the scattering of DM in the Earth and at detector, carefully modelling the Earth internal composition, and considering different target materials for the assumed directional and non-directional DM direct detection experiments.

We have found that future DM direct detection experiments with an ultra-low energy threshold  of about 1 eV (and equally low energy resolution) have the potential to reveal the population of DM particles studied in this paper with the same exposure needed to detect the associated Milky Way DM  component.~DM particles bound to the Earth manifest as a prominent feature in the low-energy part of the observed nuclear recoil energy spectrum. In particular we have found that   DM-nucleus operators like $\mathcal{O}_{11}$ can give rates in recoil events of bound DM in detectors up to a few hundred times higher than the corresponding Milky Way DM in low energies.~The existence and the shape of this feature are independent of the dark matter-nucleus scattering cross-section normalisation.~This work provides an additional important motivation to invest in the design and development of a new class of ultra-low threshold energy detectors. 
\\

{{\it Acknowledgments.}~CK is partially funded by the Danish National Research Foundation, grant number DNRF90 and by the Danish Council for Independent Research, grant number DFF – 4181-00055.

\section{Appendix}
We show what is the value of $\cos\theta_q$ in the generic case of a nonzero $\phi$. Recall that $\cos\theta_q=\hat{\theta}\cdot \hat{\ell}$ with $\hat{\theta}$ and $\hat{\ell}$ are given in Eqs.~(\ref{hattheta}) and (\ref{hatel}) (see also Eqs.~(\ref{dxl}) and (\ref{dyl})).  Note the coordinate system convention  we have used: the perihelion, the center of the Earth and the detector lie on the $x-y$ plane with the perihelion being on the $x$-axis. In order to find $\cos\theta_q$ for a nonzero value of $\phi$, we need to go to a reference system that is rotated by an angle $\phi$ around the axis that connects the center of the Earth and the detector. Practically this can be achieved by the following coordinate transformations: i) We rotate around the $z$-axis by an angle $\theta$.  This will make the $x$-axis pass through the detector. ii) We rotate around the new $x$-axis (the axis passing from the detector and the center of the Earth) by an angle $\phi$. iii) We rotate around the new $z$-axis by an angle $-\theta$. The new coordinate system will be given in terms of the old one as
\beq
\label{ccc}
\hat{\mathbf{x}}\mathbf{'}=C_3 \cdot C_2 \cdot C_1 \cdot \hat{\mathbf{ x}},
\eeq
where $\hat{\mathbf{x}}\mathbf{'}=(\hat{x}',\hat{y}',\hat{z}')$, $\hat{\mathbf{x}}=(\hat{x},\hat{y},\hat{z})$ and $C_{1,2,3}$ are the $3\times 3$ rotation matrices that correspond to the rotations i), ii), iii). In particular Eq.~(\ref{ccc}) gives explicitly
\begin{widetext}
\begin{align}
\label{xyz}
\hat{x}'&=(\cos^2\theta+\sin^2\theta \cos\phi)\hat{x}+\sin\theta \cos\theta (1-\cos\phi)\hat{y}-\sin\theta\sin\phi\hat{z} \nonumber \\
\hat{y}'&=\sin\theta \cos\theta (1-\cos\phi)\hat{x}+(\sin^2\theta+\cos^2\theta \cos\phi)\hat{y}+\cos\theta\sin\phi \hat{z} \nonumber \\
\hat{z}'&=\sin\theta \sin\phi \hat{x}-\sin\phi \cos\theta \hat{y}+\cos\phi \hat{z}.
\end{align}
\end{widetext}
The velocity of the particle that follows an elliptic orbit where the perihelion is rotated around the detector axis by $\phi$ should be given by Eq.~(\ref{hatel})
with $\hat{x}$ and $\hat{y}$ substituted by $\hat{x}'$ and $\hat{y}'$ respectively
\beq
\hat{\ell}=\frac{dx}{d\ell}\hat{x}'+\frac{dy}{d\ell}\hat{y}',
\eeq
with  $dx/d\ell$ and $dy/d\ell$ given from Eqs.~(\ref{dxl}) and (\ref{dyl}). Using Eq.~(\ref{xyz}) we  calculate
$\cos\theta_q=\hat{\theta}\cdot \hat{\ell}$ and obtain Eq.~(\ref{thetaq2}).

\end{document}